\documentclass[journal]{new-aiaa}
\usepackage{amstext}
\usepackage{textcomp}
\usepackage[utf8]{inputenc}
\usepackage{setspace}
\usepackage{graphicx}
\usepackage{longtable,tabularx,booktabs}
\usepackage{xcolor}
\usepackage{dashrule}
\usepackage{threeparttable}
\usepackage{tikz}
\usetikzlibrary{shapes}

\definecolor{magenta2}{RGB}{255,0,255}
\definecolor{blue2}{RGB}{0, 127, 255}
\definecolor{grey2}{RGB}{85, 85, 85}
\tikzstyle{longdashed}=                  [dash pattern=on 6pt off 2pt]
\tikzstyle{dashdotdot}=              [dash pattern=on 4pt off 2pt on \the\pgflinewidth off 1pt on \the\pgflinewidth off 2pt]

\title{Wall Modeling of Turbulent Flows with Varying Pressure Gradients Using Multi-Agent Reinforcement Learning \footnote{Portions of this work were presented in AIAA Paper 2023-3985, AIAA AVIATION 2023 Forum, San Diego, CA, 12--16 June 2023.}}

\author{Di Zhou \footnote{Postdoctoral Scholar Research Associate in Aerospace, Graduate Aerospace Laboratories, Member AIAA}}
\author{H. Jane Bae \footnote{Assistant professor, Graduate Aerospace Laboratories, Senior Member AIAA}}
\affil{California Institute of Technology, Pasadena, CA 91125, USA.}

\begin{document}

\maketitle

\begin{abstract}

We propose a framework for developing wall models for large-eddy simulation that is able to capture pressure-gradient effects using multi-agent reinforcement learning. Within this framework, the distributed reinforcement learning agents receive off-wall environmental states including pressure gradient and turbulence strain rate, ensuring adaptability to a wide range of flows characterized by pressure-gradient effects and separations. Based on these states, the agents determine an action to adjust the wall eddy viscosity, and consequently the wall-shear stress. The model training is \emph{in situ} with wall-modeled large-eddy simulation grid resolutions and does not rely on the instantaneous velocity fields from high-fidelity simulations. Throughout the training, the agents compute rewards from the relative error in the estimated wall-shear stress, which allows the agents to refine an optimal control policy that minimizes prediction errors. Employing this framework, wall models are trained for two distinct subgrid-scale models using low-Reynolds-number flow over periodic hills. These models are validated through simulations of flows over periodic hills at higher Reynolds numbers and flow over the Boeing Gaussian bump. The developed wall models successfully capture the acceleration and deceleration of wall-bounded turbulent flows under pressure gradients and outperform the equilibrium wall model in predicting skin friction.

\end{abstract}

\section{Nomenclature}

{\renewcommand\arraystretch{1.0}
\noindent\begin{longtable*}{@{}l @{\quad=\quad} l@{}}
$a$ & action of reinforcement learning agent \\
$B$  & intercept constant of the log law of the wall \\
$C_p$ & mean pressure coefficient \\
$C_f$ & mean skin friction coefficient \\
$f$ & geometrical function of the Boeing Gaussian bump surface \\
$H$ & height of periodic hill \\
$h$ & maximum height of the Boeing Gaussian bump \\
$h_m$ & wall-normal distance of the reinforcement learning agent \\
$L$ & width of the Boeing Gaussian bump \\
$L_x, L_y, L_z$ & dimensions of computational domain in $x$, $y$ and $z$ directions \\
$N_{\text{CV}}$ & total number of mesh cells \\
$N_x, N_y, N_z$ & number of mesh cells in $x$, $y$ and $z$ directions \\
$n$ & wall-normal direction pointing towards the interior of flow field \\
$p$ & pressure \\
$Re$ & Reynolds number \\
$r$ & reward of reinforcement learning agent \\
$S_{1}, S_2, S_3, S_4$ & The first, second, third and fourth environmental states of the developed reinforcement learning wall model \\
$S_{12}$ & turbulence shear strain rate \\
$s$ & wall-parallel direction pointing towards the positive $x$ direction \\
$t$ & dimensionless time \\
$U_b$ & bulk velocity at the top of hill \\
$U_{\infty}$ & freestream velocity \\
$u_n$ & velocity in wall-normal direction \\
$u_p$ & velocity scale based on pressure gradient \\
$u_s$ & velocity in wall-parallel direction \\
$u_x$ & velocity in $x$ direction \\
$u_y$ & velocity in $y$ direction \\
$u_\tau$ & friction velocity \\
$u_{\tau p}$ & composite friction velocity \\
$x,y,z$ & coordinates in the streamwise (freestream-aligned), vertical and spanwise directions \\
$x_{\text{sep}}$ & mean separation location \\
$x_{\text{rea}}$ & mean reattachement location \\
$y_n$ & wall-normal distance \\
$\alpha$ & action scale of reinforcement learning agent \\
$\Delta T$ & simulation time step size \\
$\Delta x, \Delta y, \Delta z$ & mesh resolutions in $x$, $y$ and $z$ direction \\
$\kappa$ & von K\'arm\'an constant \\
$\nu$ & kinematic viscosity \\
$\nu_t$ & eddy viscosity \\
$\rho$ & fluid density \\
$\tau_w$ & wall-shear stress \\
$\left|\cdot\right|$ & absolute value operator \\
$\langle \cdot \rangle$ & temporal and spanwise averaging operator \\
$\overline{(\cdot)}$ & temporal averaging operator\\

\multicolumn{2}{@{}l}{Subscripts}\\
$i$ & index of time step \\
$m$ & modeled quantity \\
$\text{rms}$ & root-mean-square value\\
$w$ & quantity defined on the wall \\
$\infty$ & freestream value \\

\multicolumn{2}{@{}l}{Superscripts}\\
$\text{ref}$ & reference value\\
$'$ & fluctuation from mean value \\
$+$ & quantity in wall unit (nondimensionalized with $u_\tau$ and $\nu$) \\
$*$	& quantity nondimensionalized with $u_{\tau p}$ and $\nu $

\end{longtable*}}

\section{Introduction}

\lettrine{L}{arge-eddy} simulation (LES) is an essential technology for the simulation of turbulent flows. The basic premise of LES is that energy-containing and dynamically important eddies must be resolved consistently throughout the domain. However, this requirement is hard to meet in the near-wall region, as the stress-producing eddies become progressively smaller. Because of the considerable cost involved in resolving the near-wall region, routine use of wall-resolved LES (WRLES) is far from being an industry standard, where short turnaround times are needed to explore high-dimensional design spaces. Consequently, most industrial computational fluid dynamics (CFD) analyses still rely on cheaper, albeit potentially less precise, lower-fidelity Reynolds-Averaged Navier-Stokes (RANS) tools. This has motivated the development of the wall-modeled LES (WMLES) approach. In WMLES, LES predicts turbulence in the outer region of the boundary layer, while a reduced-order model on a comparatively coarser grid addresses the impact of the energetic near-wall eddies. It not only significantly reduces the grid resolution requirement but also allows for a larger time step size. Because of such characteristics, WMLES has been anticipated as the next step to enable the increased use of high-fidelity LES in realistic engineering and geophysical applications. 

The most popular and well-known WMLES approach is the so-called RANS-based wall modeling \citep{balaras1996two, piomelli2002wall, wang2002, chung2009large, park2014, larsson2016large, bose2018wall}, which computes the wall-shear stress using the RANS equations. To account for the effects of nonlinear advection and pressure gradient, the unsteady three-dimensional RANS equations are solved \citep{wang2002, park2014}. However, these models assume explicitly or implicitly a particular flow state close to the wall (e.g. fully-developed turbulence in equilibrium over a flat plate) and/or rely on RANS parametrization which is manually tuned for varying pressure-gradient effects. To remove the RANS legacy in wall modeling, \citet{bose2014} and \citet{bae2019} proposed a dynamic wall model using slip wall boundary conditions for all three velocity components. Although these models are free of \emph{a priori} specified coefficients and add negligible additional cost compared to the traditional wall models, they were found to be sensitive to numerics of the flow solver and subgrid-scale (SGS) model, which hinders the application of these wall models in WMLES. The recent rise of machine learning has prompted supervised learning as an attractive tool for discovering robust wall models that automatically adjust for different conditions, such as variations in the pressure gradient. \citet{zhou2021} proposed a data-driven wall model that considers pressure-gradient effects, developed through supervised learning from WRLES data of periodic-hill channel flow. They subsequently refined this model, using a similar training strategy while integrating the law of the wall \citep{zhou2023awall}. While these trained models performed well in \emph{a priori} testing for a single time step, they broke down in \emph{a posteriori} testing due to integrated errors that could not be taken into account via supervised learning \citep{vadrot2023survey}. Moreover, supervised learning typically demands an abundance of training data that is generated from higher-fidelity simulations such as direct numerical simulation (DNS) or WRLES in the form of instantaneous flow field data. This  requirement can add additional costs to the learning process. It is worth noting that \citet{lozano2023machine} recently introduced an innovative LES wall modeling approach through supervised learning, termed the ``building-block-flow wall model''. This model is tailored to account for various flow configurations, including wall-bounded turbulence with pressure gradients and separations, by representing the flow as a combination of simple canonical flows (or building blocks). The model integrates a classifier that discerns local flow characteristics by comparing them to an array of pre-defined building-block flows, along with a predictor that estimates the wall-shear stress by integrating these building blocks. A salient feature of this model is its reliance on data sourced directly from WMLES, which not only ensures training data consistency with both the numerical discretization and the meshing strategy of the flow solver but also reduces the costs of the training process. This approach was later extended to a unified SGS and wall model, demonstrating promising results in simulating realistic flow configurations \citep{ling2022wall, arranz2023wall, lozano2023building}.

Reinforcement learning (RL) is an important machine learning paradigm with foundations on dynamic programming \citep{bib:BertsekasRL} and it has been used in the applications of flow control \citep{gazzola2014,novati2017} and SGS model development \citep{novati2021automating}. Recently, \citet{bae2022scientific} proposed a framework of wall model development based on multi-agent reinforcement learning (MARL) and demonstrated its efficacy in canonical channel and zero-pressure-gradient (ZPG) boundary layer flows. \citet{vadrot2023log} further improved the framework and trained a wall model capable of predicting the log law in channel flows across an extended range of Reynolds numbers. These studies underscore considerable potential of RL as a model development tool. In the frameworks proposed by \citet{bae2022scientific} and \citet{vadrot2023log}, a series of RL agents are distributed along the computational grid points, with each agent receiving local states and rewards and then providing local actions at each time step. The trained RL-based wall models (RLWMs) match the performance of the RANS-based equilibrium wall model (EQWM) \citep{cabot2000approximate, kawai2012}, which has been tuned for the log law of the wall. However, the RLWMs are able to achieve these results through training on moderate Reynolds number flows, guided by a reward function solely based on the recovery of the correct mean wall-shear stress. Furthermore, instead of relying on \emph{a priori} knowledge or RANS parametrization to perfectly recover the wall boundary condition computed from filtered DNS data, RL can develop novel models that are optimized to accurately reproduce the flow quantities of interest. This is achieved by discovering the dominant patterns in the flow physics, which enables the model to generalize beyond the specific conditions used for training. Therefore, the models are trained \emph{in situ} with WMLES and do not require any higher-fidelity velocity fields. 

Building upon the methodologies of \citet{bae2022scientific} and \citet{vadrot2023log}, we adapt the frameworks in the present study for turbulent flows subject to pressure gradients. Specifically, we train wall models using low-Reynolds-number flow over periodic hills with states that inform pressure-gradient effects, and subsequently test them on flows with higher Reynolds numbers and different configurations. Our first objective is to develop wall models for LES based on MARL that are robust to pressure-gradient effects in a data-efficient way. Another objective of this study is to evaluate the applicability of the trained wall models by simulating a flow configuration distinct from the one used in training. Specifically, we focus on the flow over a three-dimensional tapered Gaussian bump \citep{slotnick2019integrated}, commonly referred to as the Boeing Gaussian bump. It is a canonical case of smooth-body separation of a turbulent boundary layer (TBL) subject to pressure-gradient and surface-curvature effects. As a widely studied flow configuration, extensive experimental data exist \cite{williams2020experimental, gray2021new, gray2022experimental, gray2022benchmark, gluzman2022simplified} for validating CFD codes, and various computational approaches, including RANS methods \citep{williams2020experimental}, DNS \citep{balin2021direct}, hybrid LES–DNS \citep{wright2021unstructured,uzun2022high}, WMLES \citep{iyer2021wall,whitmore2021large,agrawal2022non} and detached-eddy simulations (DES) \cite{balin2020wall} have been evaluated and compared for the flow around the bump. The WMLES studies \citep{iyer2021wall,whitmore2021large,agrawal2022non} showed that the performance of the classical EQWM is still less than satisfactory, particularly for predicting the separation bubble and the peak value of skin friction. In the current study, the detailed comparison between simulations using the EQWM and the developed RLWMs are conducted in terms of the mean skin friction as well as the mean pressure on the bump surface, and the prediction of the velocity field. Earlier iterations of this study are documented in \citep{zhou2022RLWM, zhou2023large}.

The remainder of this paper is organized as follows. In Section~\ref{sec:model}, the details of wall-model training based on the flow over periodic hills are introduced. In addition, the validation results of the developed wall models for flow over periodic hills at the Reynolds number of training and higher Reynolds numbers are presented. Section~\ref{sec:bump} outlines the simulation setup for the flow over the Boeing Gaussian bump, and demonstrates the results of mean skin friction, mean pressure coefficients on the bump surface, and mean velocity in the flow field. Additionally, the simulations with the developed wall models are compared to those using the EQWM and experimental data. Finally, Section~\ref{sec:con} summarizes the conclusions drawn from this study.

\section{Model Development and Validation}
\label{sec:model}

\subsection{Methodology}
\label{sec:method}

\subsubsection{Flow simulation}

For the flow solver, we utilize a finite-volume, unstructured-mesh LES code \citep{you2008discrete}. The spatially-filtered incompressible Navier-Stokes equations are solved with second-order accuracy using cell-based, low-dissipative, and energy-conservative spatial discretization and a fully-implicit, fractional-step time-advancement method with the Crank–Nicolson scheme. The Poisson equation for pressure is solved using the bi-conjugate gradient stabilized method \citep{van1992bi}. In addition, to study the effect of SGS model on the training and performance of wall model, both the dynamic Smagorinsky model (DSM) \citep{germano1991dynamic,lilly1992} and the Vreman model \citep{vreman2004eddy} are used in the training and test simulations. More specifically, the model constant $c$ in the Vreman model is set to 0.025 in the present study, as \citet{zhou2024sensitivity} suggests that this value potentially offers a more robust performance in simulating separated turbulent flow. The reliability of this LES code in accurately simulating turbulent flows has been demonstrated in various configurations, such as rough-wall TBLs \cite{yang2013boundary}, flow over an axisymmetric body of revolution \cite{zhou2020large}, and rotor interactions with thick axisymmetric TBL \cite{zhou2022computational}.

For training the wall model, we prefer a flow configuration that (i) has widely available wall-shear stress profiles for several Reynolds numbers and (ii) does not require tuning of the inlet profile or other boundary conditions. The flow over periodically arranged hills in a channel as proposed by \citet{mellen2000large} has well-defined boundary conditions, can be computed at affordable costs, and nevertheless inherits all the features of a flow separating from a curved surface and reattaching on a flat plate. Furthermore, the periodic-hill channel does not require configuring the inlet boundary condition for different grid resolutions and wall models, which is necessary for non-periodic flows. This configuration has become a popular benchmark test case for validating CFD codes. Numerous experimental and high-fidelity numerical references, such as \citep{frohlich2005highly, rapp2011flow, krank2018direct, gloerfelt2019large, xiao2020flows}, exist and provide extensive data across a wide range of Reynolds numbers, spanning $700\le Re_H \le 37000$, where $Re_H$ is the Reynolds number based on hill height $H$ and bulk velocity at the top of hill $U_b$. 

The periodic-hill channel flow configuration has the dimensions of $9H\times3.035H\times4.5H$ in streamwise ($x$), vertical ($y$), and spanwise ($z$) directions, respectively. In the simulations of the present study, periodic boundary conditions are applied on streamwise and spanwise boundaries, and the EQWM is employed at the top wall. To maintain constant bulk velocity in time, the flow is driven by a time-varying body force following the control algorithm proposed by \citet{balakumar2014dns}. Two meshes with different densities are used in the present study, and the details of the meshes are listed in Table~\ref{table1}. The meshes are evenly spaced in $z$ direction, and approximately uniform in both $x$ and $y$ directions. Moreover, a maximum Courant--Friedrichs--Lewy (CFL) number of 1 is used for all simulations.

\begin{table}
\begin{center}
\vskip 0.0in
\caption{Simulation cases in comparison to reference data, including mesh size, mean separation location $x_{\text{sep}}$ and mean reattachment location $x_{\text{rea}}$ at $Re_H=10595$. \label{table1}}
\begin{tabular}{l c c c}
\toprule
Case & Mesh size ($N_x\times N_y\times N_z$) & $x_{\text{sep}}/H$ & ${x_{\text{rea}}/H}$ \\
\midrule
RLWM-DSM & $128\times64\times64$ & 0.29 & 4.57 \\
RLWM-DSM, coarse mesh & $64\times32\times32$ & 0.43 & 3.66 \\ 
RLWM-VRE & $128\times64\times64$ & 0.39 & 3.96 \\ 
RLWM-VRE, coarse mesh & $64\times32\times32$ & 0.36 & 4.68 \\
EQWM-DSM & $128\times64\times64$ & 0.57 & 3.05 \\ 
EQWM-VRE & $128\times64\times64$ & 0.58 & 3.28 \\
DNS \citep{krank2018direct} & $896\times448\times448$ & 0.20 & 4.51 \\ 
WRLES \citep{gloerfelt2019large} &  $512\times256\times256$ & -0.11 & 4.31 \\
\bottomrule
\end{tabular}
\end{center}
\end{table}


\subsubsection{Reinforcement learning architecture}

The MARL architecture of wall-model training in the present study is based on the one proposed by \citet{bae2022scientific}. During model training, the agents distributed above the wall receive states based on local instantaneous flow information and a reward based on the estimated wall-shear stress, then provide local actions to update the wall boundary condition at each time step. The agents infer a single optimized policy through their repeated interactions with the flow field to maximize their cumulative long-term rewards.

In order to utilize MARL as a tool for wall-model development, an RL toolbox, smarties \citep{novati2019a}, is coupled with the aforementioned unstructured-mesh flow solver. The RL tool is an open-source C\texttt{++} library and is optimized for high CPU-level efficiency through fine-grained multi-threading, strict control of cache-locality, and computation-communication overlap. It implements many established deep RL algorithms as well as methods that have been systematically tested for data-driven turbulence modeling \citep{novati2021automating}. In our current research, we adopt a deep RL algorithm that combines an off-policy actor-critic method, V-RACER, with the Remember-and-Forget Experience Replay (ReF-ER) algorithm \citep{novati2019a}. It can efficiently identify optimal policies within the wide action space inherent to fluid flow systems, even in situations with partial measurements and limited observability. The effective coupling between the RL tool and the flow solver has been validated using the same training and testing configurations as the study of \citet{bae2022scientific}.


\subsubsection{Training of the wall model}\label{sec:training}

The RLWM training is conducted using the LES of periodic-hill channel flow at $Re_H=10595$ with a baseline mesh ($N_x\times N_y\times N_z=128\times64\times64$). A total of 512 agents are uniformly distributed along the bottom wall, and the wall-normal locations of the agents $h_m$ are randomly selected between $0.01H$ and $0.09H$ at each agent location. Compared to fixed agent locations, this approach increases the diversity of the environmental state samples encountered during model training. Moreover, the friction Reynolds number, $Re_\tau$, for the flow on the bottom of the channel reaches maximum value of approximately 1900.

Considering the existence of log law for inner-scaled mean streamwise velocity profile ($\langle u_s \rangle ^+=(1/\kappa)\ln{y_n^+}+B$) in the near-wall region of turbulent flows, \citet{bae2022scientific} developed a wall model for flat-plate channel flow using two instantaneous flow states based on the log law: $(\partial{u_s}/\partial{n})^+h_m^+$ and $u_s^+-(\partial{u_s}/\partial{n})^+h_m^+\ln{(h_m^+)}$. Note that the inner scaling of the instantaneous quantities for the training uses the modeled $u_\tau$ rather than the true $u_\tau$, without relying on any empirical coefficients. Recently, \citet{vadrot2023log} improved the performance and effectiveness of the model at different Reynolds numbers by using $\left[(\partial{u_s}/\partial{n})^+h_m^+-1/\kappa^{\text{ref}}\right]\ln{(h_m^+)}$ and $u_s^+-(\partial{u_s}/\partial{n})^+h_m^+\ln{(h_m^+)}$ as model states, where $\kappa^{\text{ref}}$ is a reference von K\'arm\'an constant. On the foundation of these studies, for the current RLWMs, we set the local instantaneous flow quantities
\begin{equation}  S_1=\left[\left(\frac{\partial{u_s}}{\partial{n}}\right)^*h_m^*-\frac{1}{\kappa^{\text{ref}}}\right]\ln{(h_m^*)} \qquad \textrm{and} \qquad S_2=u_s^*-\left(\frac{\partial{u_s}}{\partial{n}}\right)^*h_m^*\ln{(h_m^*)}
\end{equation}
as the first two model states, where $\kappa^{\text{ref}}=0.41$. Our objective with the MARL framework is to develop a wall model that adapts to varying pressure gradients and accurately captures flow separation. Ideally, this wall model should be capable of recognizing different flow regimes (e.g., those with favorable or adverse pressure gradients) based on local instantaneous flow states, then providing appropriate action accordingly. To enhance adaptability for flows with varying pressure gradients and separations, we have incorporated the turbulence strain rate and the local wall-parallel pressure gradient parameter: 
\begin{equation} S_3=S^*_{12}= \frac{1}{2}\left[\left(\frac{\partial{u_s}}{\partial{n}}\right)^*+\left(\frac{\partial{u_n}}{\partial{s}}\right)^*\right] \qquad \textrm{and} \qquad S_4=\left(\frac{\partial{p}}{\partial{s}}\right)^*h_m^*
\end{equation}
as the third and the fourth model states, respectively. Furthermore, to increase the applicability of the wall model for a wide range of flow parameters, the states are nondimensionalized using kinematic viscosity $\nu$ and the composite friction velocity $u_{\tau p}=({u_{\tau}^2+u_{p}^2})^{1/2}$ introduced by \citet{manhart2008}, where $u_p=|(\nu/\rho)( \partial{p_w}/\partial{s})|^{1/3}$, $p_w$ is the pressure on the bottom wall and $u_{\tau}$ is the friction velocity based on the modeled wall-shear stress.  

It is well known that the approach to applying wall stress model in WMLES affects the prediction of mean flow quantities \citep{bae2021effect}. A previous study \citep{zhou2022RLWM} tested two formulations of the boundary condition, namely the wall-shear-stress and the wall eddy-viscosity formulations, to determine the more appropriate action for the wall model. It was found that the wall eddy-viscosity formulation offers greater robustness for WMLES of separated flows, leading us to adopt this approach for the current wall model. Specifically, in the model, each agent acts to adjust the local wall eddy viscosity $\nu_{t,w}$ at each time step through a multiplication factor $\nu_{t,w}(t_{i+1}) = a\nu_{t,w}(t_i)$, where $a\in \left[1-\alpha\Delta T U_b/\Delta x,1+\alpha\Delta T U_b/\Delta x\right]$ and $\alpha$ is selected to be $10^{-3}$. The local wall-shear stress can be calculated by the formula $\tau_{w}=\rho\nu (1+\nu_{t,w}^{*})(\partial{u_s}/\partial{n})_w$. It should be noted that \citet{vadrot2023log} recommended limiting the action range for adjusting wall-shear stress, which is helpful in alleviating log-layer mismatch (LLM). Additionally, the reward $r$ is calculated based on $r(t_{i})=[|\tau_w^{\text{ref}}-\tau_w(t_{i-1})|-|\tau_w^{\text{ref}}-\tau_w(t_{i})|]/\tau_{w,\text{rms}}^{\text{ref}}$ at each location, where $\tau_w^{\text{ref}}$ and $\tau_{w,\text{rms}}^{\text{ref}}$ are the mean and root-mean-square wall-shear stress from the reference high-fidelity simulation. The reward is proportional to the improvement in the modeled wall-shear stress compared to the one obtained in the previous time step, and an extra reward of 0.1 is added when the modeled $\tau_w$ is within $10\%$ of the reference value. 

During model training, each episode (one simulation) is initialized with a normalized wall eddy viscosity, $\nu_{t,w}^*$, randomly chosen from $\left(0,10\right]$. Furthermore, the process of randomly selecting the wall-normal locations for all RL agents is repeated at the beginning of each episode. To generate the initial condition for training, the simulation is started from a flow field generated by the EQWM and run with the given initial $\nu_{t,w}^*$ for 20 flow-through times (FTTs) to remove numerical artifacts. Each episode of the model training is then conducted for 5 FTTs. Throughout these 5 FTTs, RL agents collect local instantaneous flow states and the corresponding rewards at every 100 time-step interval, while concurrently updating the local $\nu_{t,w}^*$ based on the action of each agent. The model training is advanced for 200 episodes with approximately 1.6 million policy gradient steps. Throughout these 200 episodes, the extensive interaction with the flow field allows the RL agents to progressively learn and refine a single optimized control policy for adjusting $\nu_{t,w}^*$, with the goal of maximizing their cumulative rewards. Specifically, the learning process utilizes a neural network architecture with identical parameters to the one described in the study of \citet{bae2022scientific}, featuring two hidden layers, each with 128 units and employing the Softsign activation function. More details regarding the history of the cumulative rewards and the convergence of model training are introduced in Appendix I. Besides, training a single RLWM based on the periodic-hill channel flow at $Re_H=10595$ with the baseline mesh requires approximately 51,000 core hours.

Previous studies \citep{LOZANODURAN2019532, rezaeiravesh2019systematic, whitmore2021large, iyer2022wall, agrawal2022non, zhou2024sensitivity} have highlighted the substantial impact of SGS model on WMLES, particularly in separated turbulent flows. To investigate the influence of the SGS model on RLWM training outcomes, we utilized two distinct SGS models namely, the DSM and the Vreman model. As detailed in Table~\ref{table1}, the model trained using the DSM is denoted as RLWM-DSM, while the one trained with the Vreman model is labeled RLWM-VRE.

\subsection{Validation}
\label{sec:results}

\subsubsection{Testing for flow over periodic hills at $Re_{H}=10595$}

To evaluate the performance of the trained RLWMs, a series of simulations for the periodic-hill channel flow at $Re_H=10595$ are carried out using meshes with different resolutions. The details of the simulation cases are listed in Table~\ref{table1}. Specifically, for the baseline mesh ($N_x\times N_y\times N_z=128\times64\times64$), also used in model training, the maximum cell sizes in the streamwise, wall-normal, and spanwise directions are 98, 52, and 93 wall units, respectively, based on the mean wall-shear stress on the bottom wall from reference DNS \citep{krank2018direct}. For the coarse mesh ($N_x\times N_y\times N_z=64\times32\times32$), these dimensions increase to maximum values of 197, 104, and 185 wall units, respectively. Moreover, the number of agents above the bottom wall is consistent with the number of mesh cells on the wall. The RL agents are located at the upper surface of wall-adjacent mesh cells in the simulations, and this placement is similar to the one adopted by \citet{vadrot2023log}. The wall eddy viscosity $\nu_{t,w}$ is updated based on the model action at every time step. All simulations are run for about 50 FTTs after initial transients. The flow statistics of all simulations are averaged over spanwise direction and time. For comparison, two baseline-mesh simulations using the EQWM that employs the traditional wall-shear stress boundary condition are carried out, one with the DSM and the other with the Vreman model. In these EQWM simulations, the center of the second off-wall cell is selected as the matching location. Additionally, the results from two high-fidelity simulations for this flow \citep{krank2018direct,gloerfelt2019large} are included as references.
  
\begin{figure}
\centering
\includegraphics[width=\textwidth]{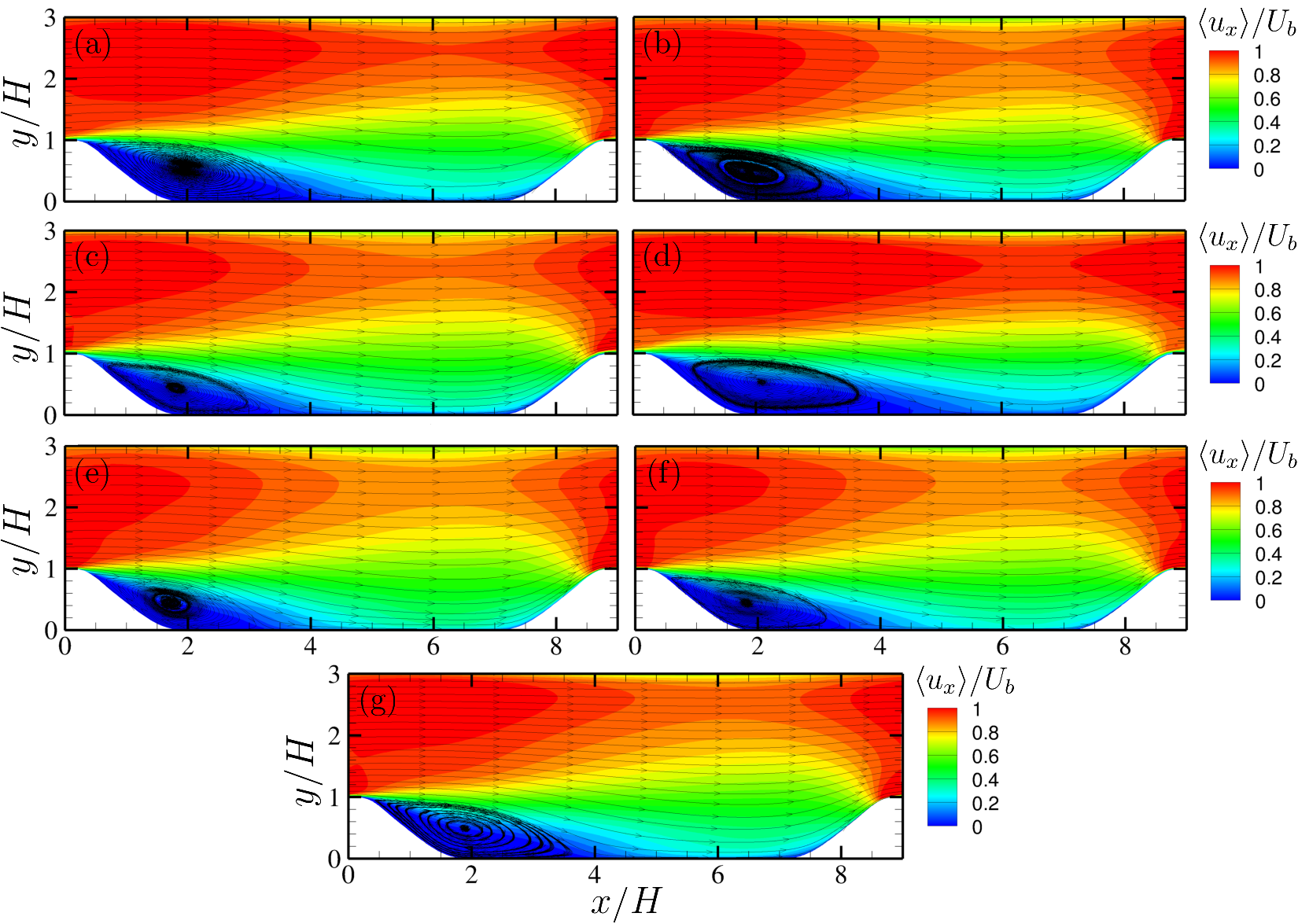}
\vspace{-0.6cm}
\caption{Contours of the mean velocity in $x$ direction and the streamlines at $Re_H=10595$ from the simulations of (a) RLWM-DSM, (b) RLWM-VRE, (c) RLWM-DSM, coarse mesh, (d) RLWM-VRE, coarse mesh, (e) EQWM-DSM, (f) EQWM-VRE, (g) WRLES \citep{gloerfelt2019large}.}
\label{fig:streamline_Re10595}
\end{figure}

Figure~\ref{fig:streamline_Re10595} shows the contours of the mean velocity in $x$ direction and the mean-flow streamlines. The flow separates on the leeward side of the hill due to a strong adverse pressure gradient (APG), and a shear layer is generated near the top of the hill. The flow reattaches in the middle section of the channel, and as the flow approaches the windward side of the downstream hill, it is subjected to a strong favorable pressure gradient (FPG) and accelerates rapidly. The simulations with the RLWMs successfully capture the separation bubble on the leeward side of the hill and yield more accurate results than the EQWM (see Table~\ref{table1} for quantitative comparison).

\begin{figure}
\centering
\includegraphics[width=0.93\textwidth]{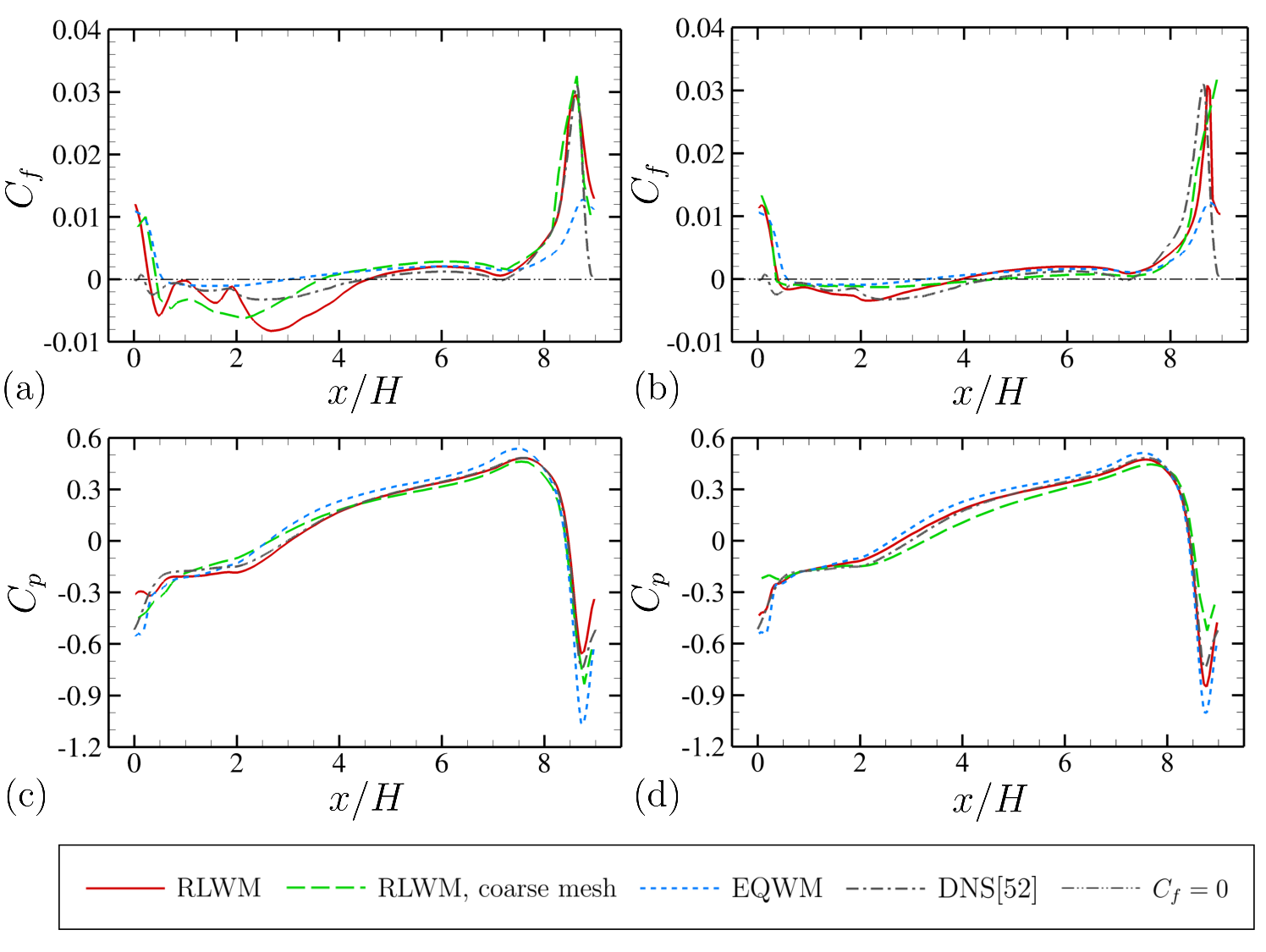}
\vspace{-0.2cm}
\caption{Comparison of (a,b) mean skin friction coefficient and (c,d) mean pressure coefficient along the bottom wall at $Re_H=10595$. The left column (a,c) presents WMLES with the DSM, while the right column (b,d) features cases with the Vreman model.}
\label{fig:CfCp_Re10595}
\end{figure}

The predictions of the mean skin friction coefficient $C_f$ and the mean pressure coefficient $C_p$ are shown in Fig.~\ref{fig:CfCp_Re10595}. The mean skin friction coefficient is defined as $C_f=\langle \tau_w\rangle/(0.5\rho U_b^2)$, where the positive direction of $\tau_w$ points toward the opposite direction of bulk flow. The mean pressure coefficient is defined as $C_p=(\langle p_w\rangle-\langle p^{\text{ref}}\rangle)/(0.5\rho U_b^2)$, where the pressure at $x/H=0$ on the top wall is chosen as reference pressure $p^{\text{ref}}$ \citep{krank2018direct}. The $C_f$ and $C_p$ are evaluated at the center of the wall boundary for each wall-adjacent control volume. Regarding $C_f$, the results from the RLWM simulations are in reasonable agreement with the DNS data, with large deviations found only near the top of the hill on the leeward side where the skin friction rapidly decreases from its maximum value to a negative value. However, the results are better than the EQWM simulations which largely under-predict the skin friction on the windward side of the hill. Furthermore, the mean locations of the separation and reattachment points (listed in Table~\ref{table1}) are better predicted by the RLWMs, consistent with the streamline shown in Fig.~\ref{fig:streamline_Re10595}. All simulations capture the qualitative trend of the mean $C_p$ on the bottom wall including the APG and FPG regimes, but large deviations among the simulation cases are visible near the top of the hill ($x/H\geq8.5$ or $x/H\leq0.5$) where the pressure sees a sudden change from strong FPG to strong APG and the flow separation emerges. Overall, the RLWMs provide more accurate predictions of $C_f$ and $C_p$ than the EQWM. When comparing the two developed wall models, it is evident that their predictions differ. Specifically, noticeable variations are seen in the results for $C_f$ and $C_p$, particularly within the regions of the separation bubble and the windward side of the hill. Additionally, the performance of these models displays a sensitivity to mesh resolution. This sensitivity is evident not only in $C_f$ and $C_p$ but also in the velocity field, as depicted in Fig.~\ref{fig:streamline_Re10595}. It should be mentioned that the environmental states used for the developed wall models do not contain any information related to the mesh resolution, which could potentially contribute to the sensitivity.

\begin{figure}
\centering
\includegraphics[width=\textwidth]{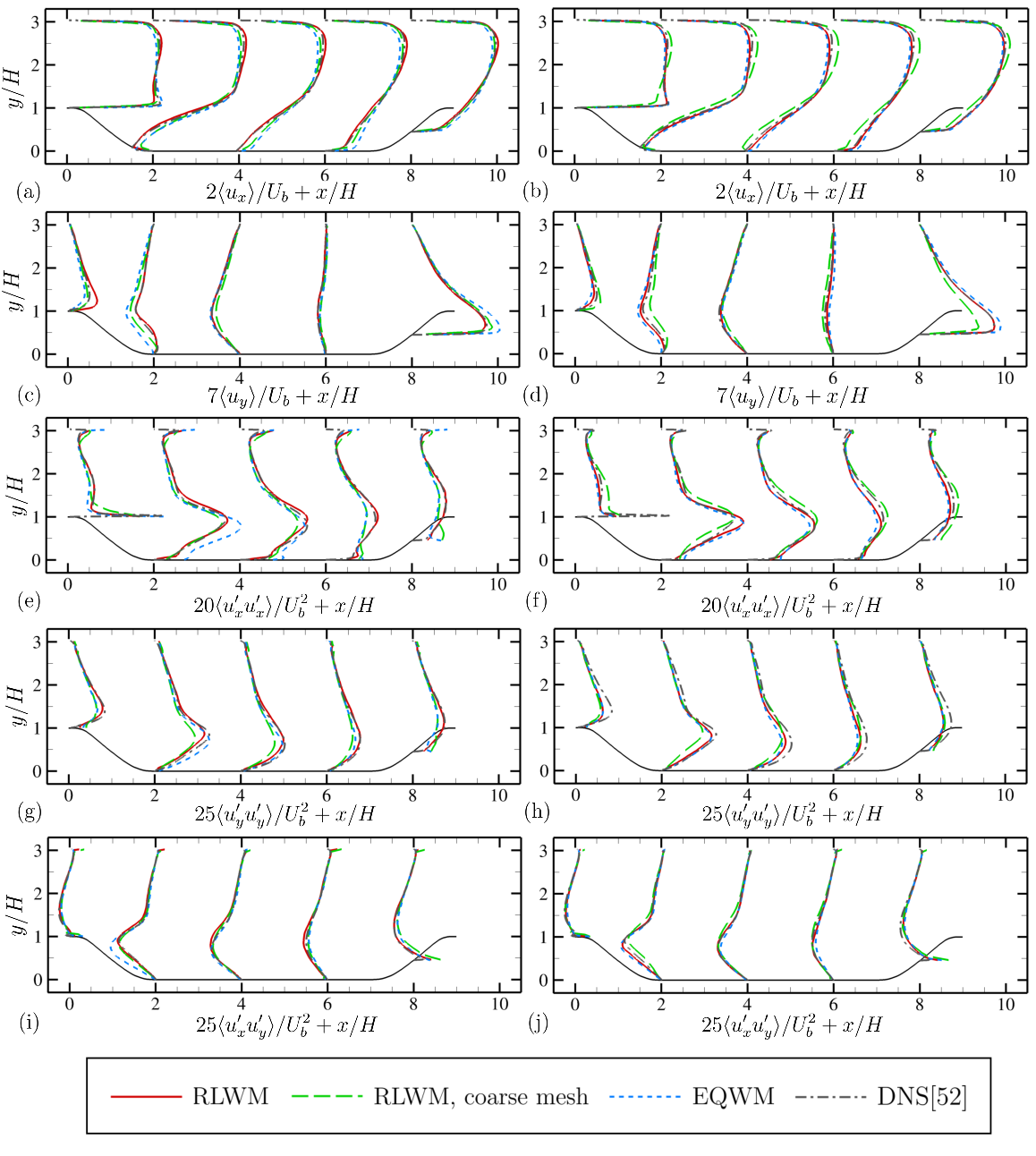}
\vspace{-0.6cm}
\caption{Comparison of (a,b) mean streamwise velocity, (c,d) mean vertical velocity, (e,f) streamwise Reynolds stress, (g,h) vertical Reynolds stress and (i,j) Reynolds shear stress at $Re_H=10595$. The left column (a,c,e,g,i) presents WMLES with the DSM, while the right column (b,d,f,h,j) features cases with the Vreman model.}
\label{fig:flowstates_Re10595}
\end{figure}

Quantitative comparison of mean velocity and Reynolds stress components at five streamwise locations ($x/H=0.05, 2, 4, 6 \text{ and } 8$) are shown in Fig.~\ref{fig:flowstates_Re10595}. The mean velocity and Reynolds stress profiles from the baseline-mesh RLWM simulations align closely with the reference DNS data. However, discrepancies are visible for the locations on the hill where the pressure gradient is strong. The EQWM simulations do not match up as well to the developed wall models, especially regarding the Reynolds stress components. Moreover, the predictions from the coarse-mesh simulations using both RLWM-DSM and RLWM-VRE are less ideal when compared to their baseline-mesh counterparts. It should be mentioned that the prediction of the velocity field not only depends on the wall boundary conditions but also on the SGS model, and the inconsistent results of RLWM simulations across different mesh resolutions may also be influenced by the varying performance of the SGS model in each mesh. 

\begin{figure}
\centering
\includegraphics[width=\textwidth]{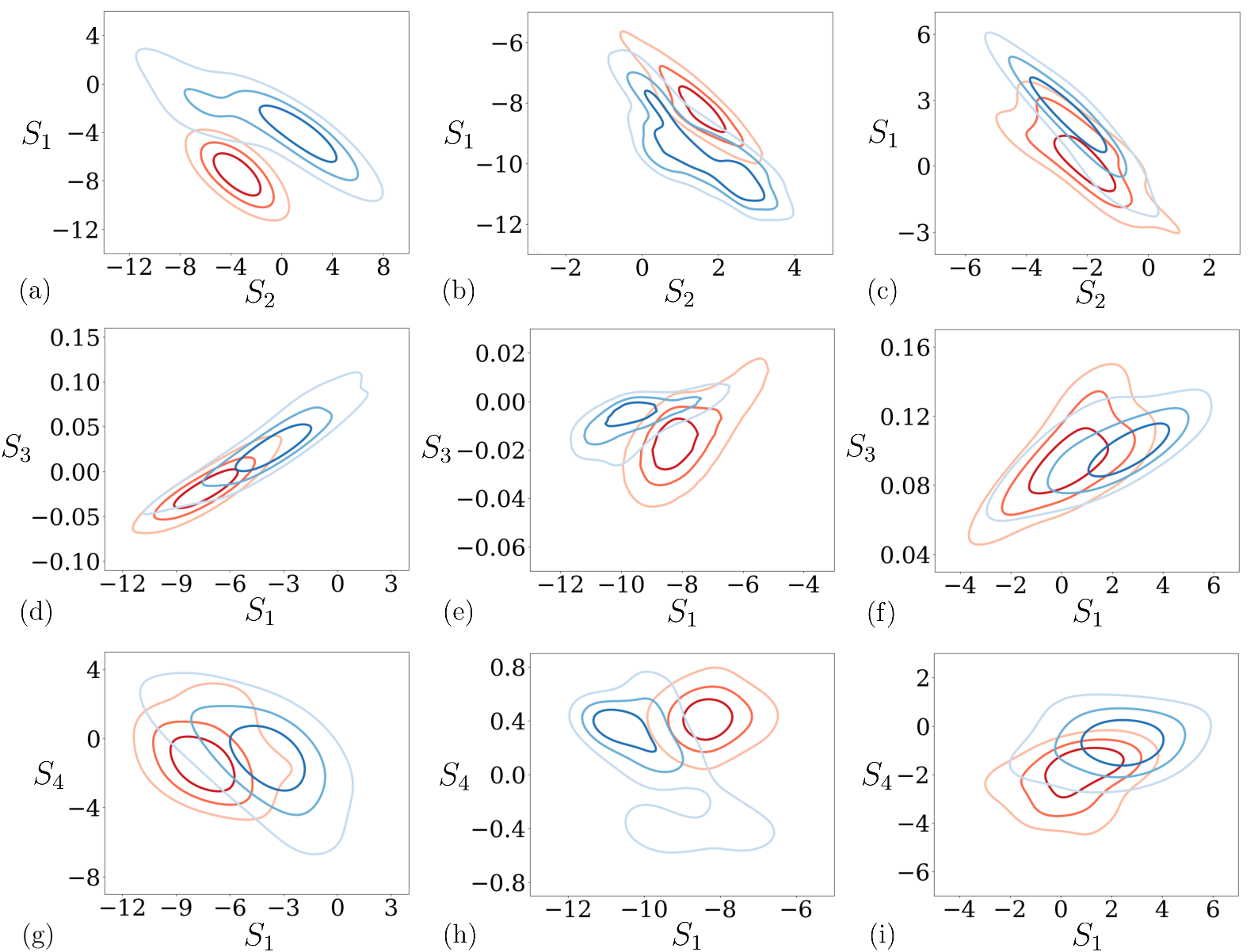}
\caption{PDFs of states for RLWM-DSM conditioned to events with $r>0.1$ and $a<0.9995$ (blue) or $a>1.0005$ (red) at (a,d,g) $x/H=0.1$, (b,e,h) $x/H=2$ and (c,f,i) $x/H=8.3$ from the baseline-mesh simulation at $Re_H = 10595$: (a--c) $S_2$ and $S_1$, (d--f) $S_1$ and $S_3$ and (g--i) $S_1$ and $S_4$. Contour levels are $25$, $50$, $75\%$ of the maximum value.}
\label{fig:state_act1}
\end{figure}

\begin{figure}
\centering
\includegraphics[width=\textwidth]{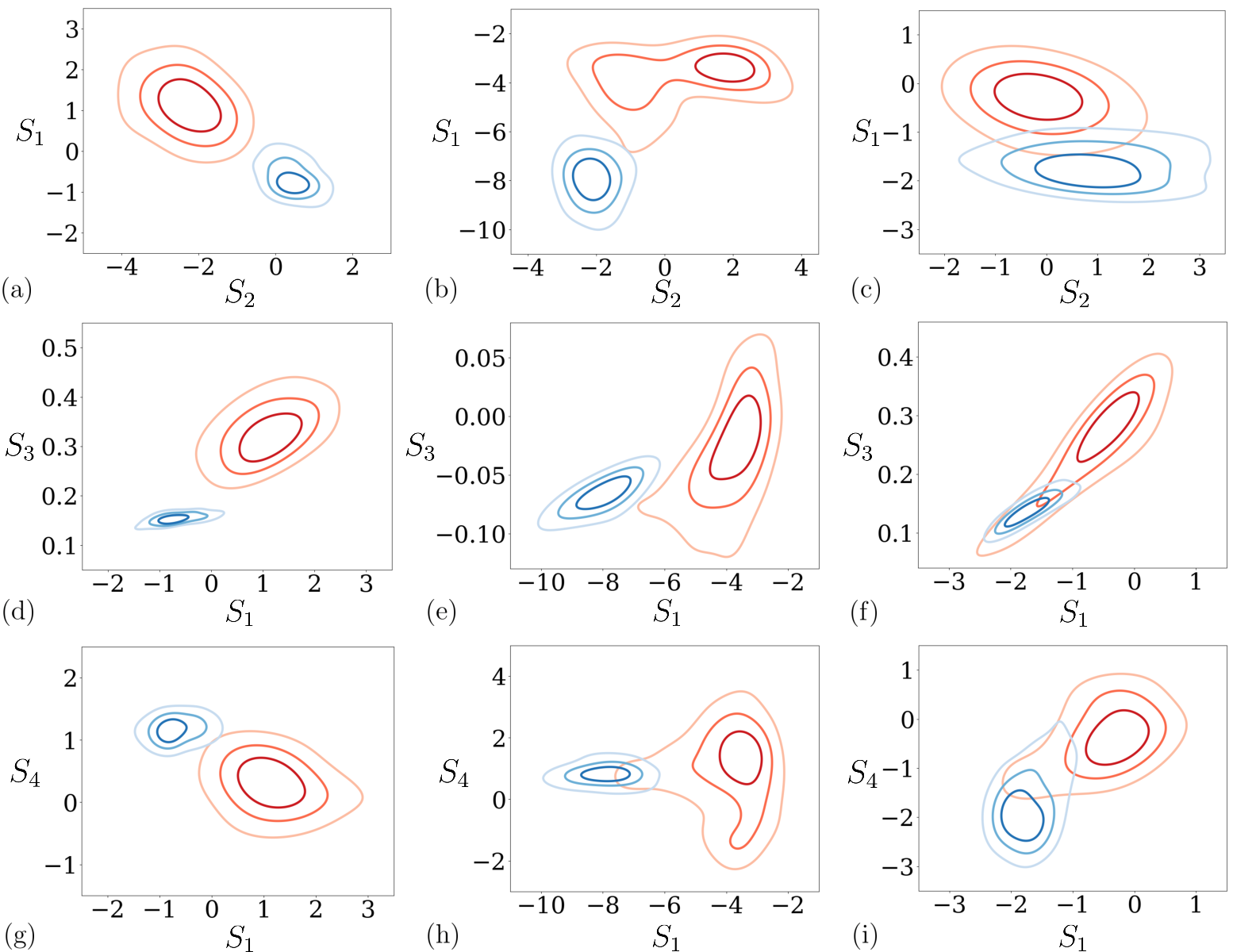}
\caption{PDFs of states for RLWM-VRE conditioned to events with $r>0.1$ and $a<0.9995$ (blue) or $a>1.0005$ (red) at (a,d,g) $x/H=0.1$, (b,e,h) $x/H=2$ and (c,f,i) $x/H=8.3$ from the baseline-mesh simulation at $Re_H = 10595$: (a--c) $S_2$ and $S_1$, (d--f) $S_1$ and $S_3$ and (g--i) $S_1$ and $S_4$. Contour levels are $25$, $50$, $75\%$ of the maximum value.}
\label{fig:state_act2}
\end{figure}

To better understand the mechanism of the trained models, we examine the state-action maps obtained from the baseline-mesh test simulations, which are the probability density functions (PDFs) of the likelihood that the models take a particular action conditioned on the occurrence of positive rewards. Such state-action maps could offer valuable insights that aid the development of empirical models for turbulent flows subjected to varying pressure gradients. Figures~\ref{fig:state_act1} and \ref{fig:state_act2} show the maps for RLWM-DSM as well as RLWM-VRE based on the distributions of instantaneous states and actions of the RL agents at three streamwise positions $x/H=0.1, 2 \text{ and } 8.5$ which are located near the top of the hill on its leeward side, within the separation bubble, and the windward side of the hill, respectively. It is worth noting that while rewards are computed for illustrative purposes in this context, they are not employed during the test simulations. Overall, the action contour lines for increasing and decreasing $\nu_{t,w}$ are well separated, which illustrates the models are able to distinguish flow states and provide appropriate actions. In particular, it is evident that RLWM-VRE offers clearer separation in the contour lines of these state-action maps than RLWM-DSM does. This suggests better performance by RLWM-VRE in the baseline-mesh simulations at $Re_H=10595$, especially in areas near the top of the hill on its leeward side and within the separation bubble (as seen in Figs.~\ref{fig:CfCp_Re10595} and \ref{fig:flowstates_Re10595}). Additionally, some discrepancies can be observed in the state-action maps between the simulations with RLWM-DSM and RLWM-VRE, particularly regarding the range of instantaneous flow states and the trends in actions. It is important to note that different SGS models are employed in each test simulation, leading to variations in the flow field, especially in the near-wall region. The impact of SGS model on flow field in WMLES has been found to be significant for non-equilibrium turbulent flows \citep{iyer2022wall, agrawal2022non, zhou2024sensitivity}. Given that the behavior of the developed RLWMs relies on the information of near-wall flow field, these variations lead to the observed discrepancies in the state-action maps. Furthermore, the RLWMs are trained using simulations that employ different SGS models, leading to variations in their control policies, which also contributes to the observed discrepancies. However, this comparison underscores the importance of considering the coupling effect between SGS model and wall model during both the training and testing phases of wall model development.

In addition, it has been shown that history effects of the pressure gradient could play a significant role in non-equilibrium wall-bounded turbulent flows \citep{bobke2017history}. The developed RLWMs implicitly account for these effects through their design and training process. Specifically, during the model training, each episode is conducted for 5 FTTs, enabling RL agents at various locations to collect local instantaneous flow states and corresponding rewards over time. This approach ensures that a degree of historical information of flow states is embedded within the training dataset. Moreover, through repeated interactions with the flow field across these episodes, the agents deduce a single optimized policy to maximize their cumulative long-term rewards. This process inherently incorporates historical data, as the decisions of agents are influenced by the outcomes of their previous actions on the flow dynamics. Additionally, in test simulations, the RLWM operates by dynamically adjusting wall eddy viscosity in response to local instantaneous flow states. This real-time adaptation, informed by temporally varying flow states, naturally integrates the history effects into the decision-making process of RLWMs.

\subsubsection{Testing for flow over periodic hills at higher Reynolds numbers}

In this section, the RLWMs are applied to WMLES of periodic-hill channel flow at $Re_H=19000$ and $37000$. The $Re_\tau$ for the flow on the bottom of the channel reaches maximum value of approximately 3300 and 5400, respectively. The simulations are conducted by using the baseline mesh ($128\times64\times64$) and the coarse mesh ($64\times32\times32$), and the implementation of the RLWMs is similar to the simulations at $Re_H=10595$. All simulations are run for about 50 FTTs after initial transients. The results from the EQWM with baseline mesh and the WRLES \citep{gloerfelt2019large} are included for comparison. 

\begin{figure}
\centering
\includegraphics[width=0.93\textwidth]{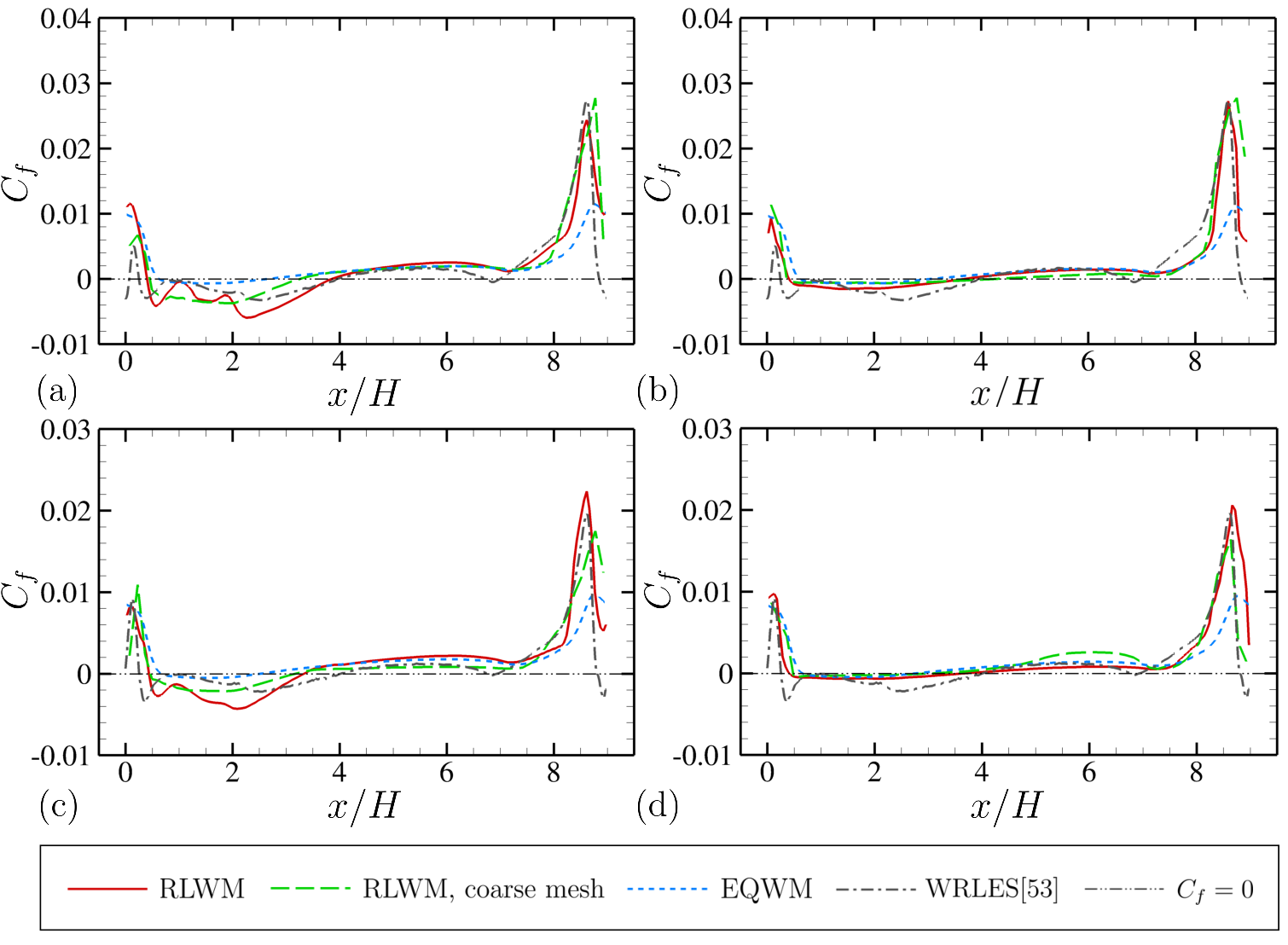}
\vspace{-0.1cm}
\caption{Comparison of mean skin friction coefficient along the bottom wall at (a,b) $Re_H=19000$ and (c,d) $Re_H=37000$. The left column (a,c) presents WMLES with the DSM, while the right column (b,d) features cases with the Vreman model.}
\label{fig:Cf_higherRe}
\end{figure}

The mean skin friction coefficients along the bottom wall at $Re_H=19000$ and $37000$ are shown in Fig.~\ref{fig:Cf_higherRe}. The distributions of $C_f$ at higher Reynolds numbers have a similar shape as the one shown in Fig.~\ref{fig:CfCp_Re10595}(a,b). As the Reynolds number increases, the peak value of the $C_f$ on the windward side of the hill decreases. The RLWM simulations align more closely with the WRLES results on hill windward side than the EQWM simulations do. Regarding the separation point, the predicted locations from the RLWM simulations are further downstream than that of the WRLES. Moreover, the RLWMs predict a reattachment location further upstream. Unlike the EQWM simulations, which considerably underestimate the size of the separation bubble, the RLWM simulations are more accurate. Similar to the aforementioned results observed at $Re_H=10595$, the RLWM predictions at these higher Reynolds numbers display sensitivity to mesh resolutions. 

\begin{figure}
\centering
\includegraphics[width=\textwidth]{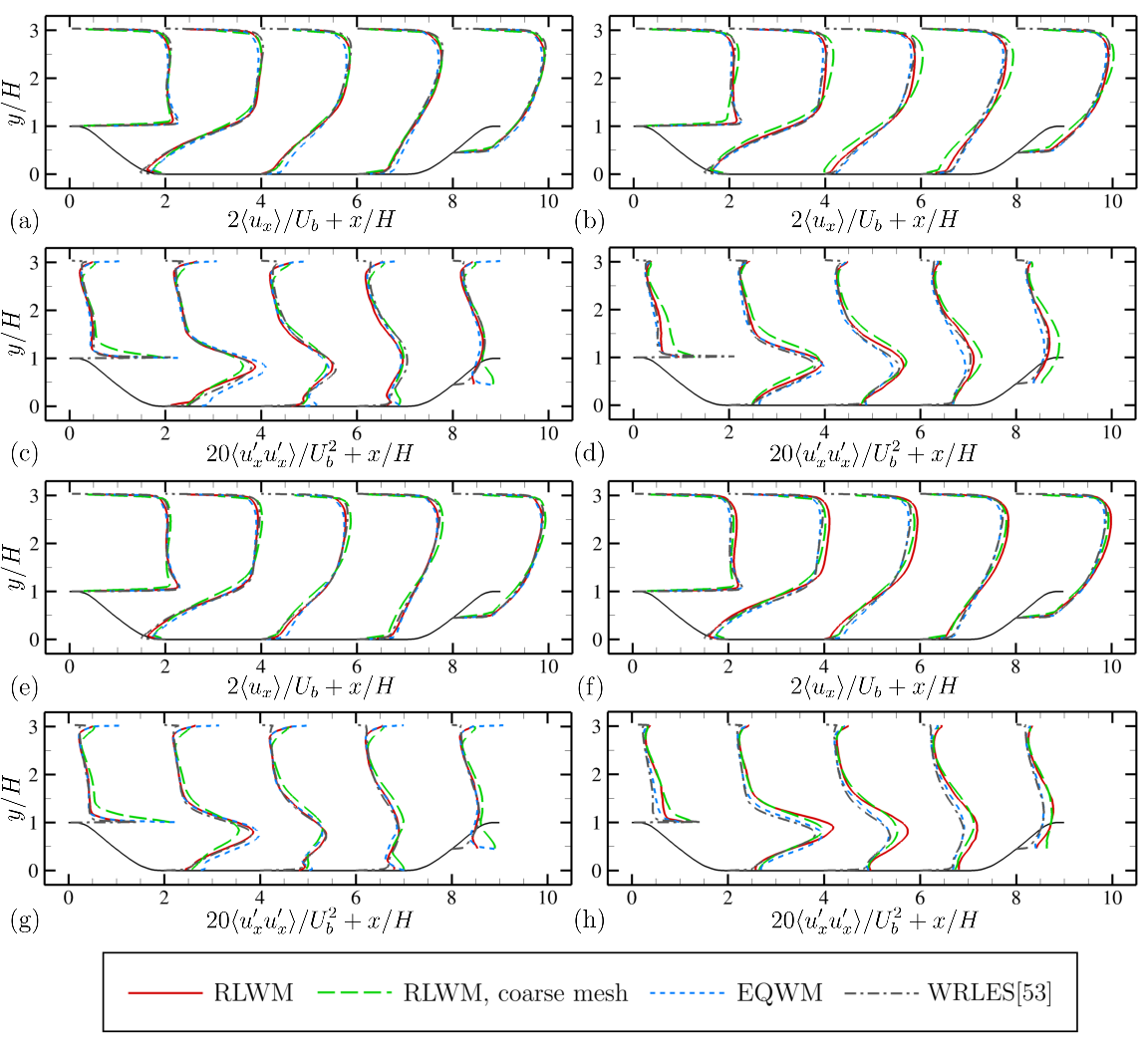}
\vspace{-0.5cm}
\caption{Comparison of (a,b,e,f) mean streamwise velocity and (c,d,g,h) streamwise Reynolds stress at (a--d) $Re_H=19000$ and (e--h) $Re_H=37000$. The left column (a,c,e,g) presents WMLES with the DSM, while the right column (b,d,f,h) features cases with the Vreman model.}
\label{fig:flowstates_higherRe}
\end{figure}

Figure~\ref{fig:flowstates_higherRe} presents the profiles of the streamwise components for both mean velocity and Reynolds stress at five streamwise stations ($x/H=0.05, 2, 4, 6, \text{ and } 8$) for $Re_H=19000$ and $37000$. As the Reynolds number increases, the discrepancies between the RLWM simulations and the WRLES profiles become more pronounced, suggesting a diminishing efficacy in the RLWMs. Among the simulations employing the developed wall models, those with RLWM-DSM yield better velocity-field predictions compared to those with RLWM-VRE, especially at $Re_H=37000$. Furthermore, when the Reynolds number rises to $Re_H=37000$, the EQWM simulations deliver velocity-field results that match or even surpass the RLWM simulations, despite their inaccurate predictions of $C_f$, as illustrated in Fig.~\ref{fig:Cf_higherRe}(c,d). 

Overall, the test simulations of periodic-hill channel flows demonstrate that the trained RLWMs are viable for handling spatially and temporally varying pressure gradients and for use at higher Reynolds numbers. However, these test simulations are limited to the same geometrical configuration as that used during training. To further assess the applicability of the trained wall models across different geometrical setups, we conduct preliminary test simulations for flat-plate channel flow at different Reynolds numbers. For the sake of conciseness, the results of these simulations are not described here in detail but are presented in Appendix II. Additionally, we also carry out test simulations for flow over the Boeing Gaussian bump, which will be the focus of the next section.

\section{Testing: Flow over Boeing Gaussian Bump}
\label{sec:bump}

\begin{figure}
\begin{center}
\includegraphics[width=0.65\textwidth]{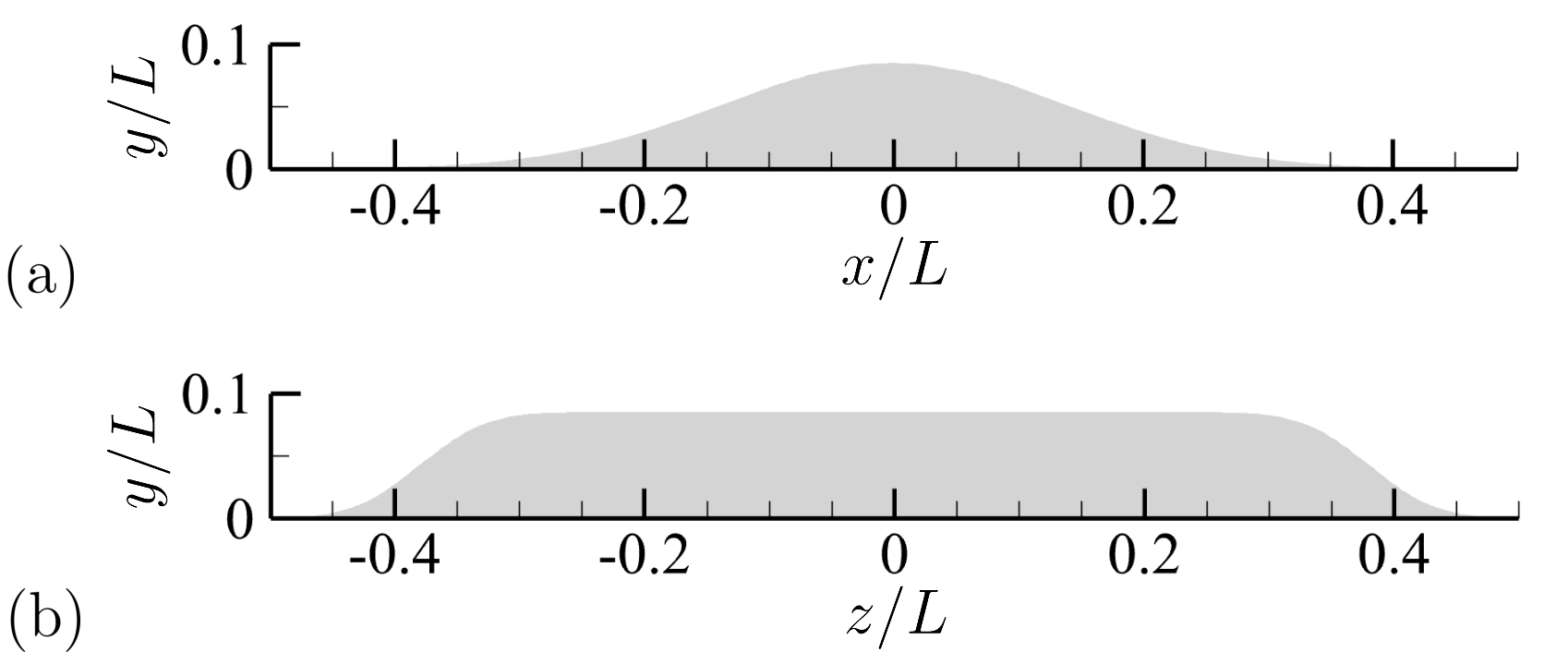}
\vspace{-0.1em}
\caption{Cross-sections of the Boeing Gaussian bump geometry showing (a) a slice along the centerline ($z/L=0$) and (b) a slice along the span ($x/L=0$).}
\label{fig:bump_xsection}
\end{center}
\end{figure}

The geometry of the Boeing Gaussian bump is given by the analytic function
\begin{equation}
    y=f(x,z) = \frac{h}{2}e^{-\left(x/x_0\right)^2}\left\{1+{\rm erf}\left[\left(\frac{L}{2}-2z_0-|z|\right)/z_0\right]\right\}~,
\end{equation}
where $h=0.085L$, $x_0=0.195L$, and $z_0=0.06L$. The cross-sections of the bump are shown in Fig.~\ref{fig:bump_xsection}. Additionally, the length scale $L$ is used to define the Reynolds number $Re_L = U_\infty L / \nu$. This geometry has been the subject of extensive experimental research in recent years \citep{williams2020experimental,gray2021new,gray2022experimental,gray2022benchmark,gluzman2022simplified}. The experimental geometry includes side and top walls, as the Gaussian bump is wall-mounted on a splitter plate inside of a wind tunnel. In this work, this geometry with side and top walls is considered. Particularly, the case of $Re_L=2\times10^6$ is simulated, and the WMLES results are compared with the experimental measurements. Experimental measurements \citep{gray2021new,gray2022experimental,gray2022benchmark,gluzman2022simplified} indicate that, under the condition of $Re_L=2\times10^6$, a maximum $Re_\tau$ of around 2000 is observed in the region ahead of the bump peak. Downstream of the separation bubble, $Re_\tau$ experiences a rapid increase, reaching an approximate value of 7500 at $x/L=0.5$, with a further increase observed at downstream locations. The range of $Re_\tau$ is beyond what was encountered during the training of the RLWMs.

\subsection{Computational details}

The present simulations employ a rectangular computational domain with the dimension of $L_x\times L_y\times L_z=2.5L\times 0.5L\times 0.5L$, which has the same blockage ratio as in the wind tunnel experiment \citep{gray2021new,gray2022experimental,gray2022benchmark,gluzman2022simplified}. The origin of the coordinate system in the domain locates at the base of the bump peak as shown in Fig.~\ref{fig:bump_xsection}. Because symmetry exists with respect to the center plane at $z/L=0$ in the geometry, the simulation domain only covers half of the entire bump span with a symmetry boundary condition applied at $z/L=0$. Besides the developed RLWMs at the bottom wall, we set the simulations with a plug flow inlet at $x/L=-1$. The side at $z/L=0.5$ and top boundary conditions at $y/L=0.5$ are treated as inviscid walls to approximate the wind tunnel condition. The outlet is placed at $x/L=1.5$ with a convective outflow boundary condition. Note that the experimental measurements \cite{gray2021new,gray2022experimental,gray2022benchmark,gluzman2022simplified} indicate the flow on the leeward side of the bump is largely unaffected by variations in the incoming TBL thickness. Previous WMLES studies \citep{whitmore2021large, agrawal2022non} also suggest the insensitivity to inlet and tunnel wall boundary conditions. The simulations are conducted using the aforementioned unstructured-mesh, incompressible, finite-volume flow solver that is coupled with the RL toolbox, smarties \citep{novati2019a}. For consistency, both the DSM and the Vreman model, previously utilized in the wall-model training, are adopted in separate simulations. Apart from the developed RLWMs, we employ the EQWM at the bottom wall in the benchmark simulations for comparison, and the center of the second off-wall cell is selected as the matching location in these simulations. Additionally, a maximum CFL number of 2 is used in all simulations. 

To study the effect of mesh resolution on the simulation results, three computational meshes with increasing resolutions in each direction are considered. These meshes consist of structured-mesh blocks covering the entire bump surface and the flat wall surfaces in both the upstream and downstream, and unstructured-mesh blocks elsewhere. The wall-normal dimension of the structured-mesh blocks is equal to $0.12L$, which covers the entire TBL on the bottom wall. Furthermore, uniform mesh resolutions are used within the structured-mesh blocks in both streamwise and spanwise directions. Along the wall-normal direction, the mesh size gradually coarsens away from the wall with an approximate stretching ratio of 1.013. Referring to the TBL thickness at $x/L=-0.683$ from the experiment \citep{gray2021new,gray2022experimental,gray2022benchmark,gluzman2022simplified}, the TBL is approximately resolved by 3 cells in the coarse mesh, 6 cells in the medium mesh, and 9 cells in the fine mesh, respectively. For the unstructured-mesh blocks, the outermost unstructured mesh has the resolution of $0.03L$ and the control volumes are refined gradually towards the bottom wall. More parameters of the computational meshes are provided in Table~\ref{tab:3d_grid}.

In the simulations using the developed RLWMs, the number of agents above the bottom surface is consistent with the number of wall cells in each mesh. Similar to the aforementioned test simulations of periodic-hill channel flows, the RL agents are positioned at the upper surface of the wall-adjacent mesh cells in the current simulations. The wall eddy viscosity $\nu_{t,w}$ is updated at each time step based on the model action. To eliminate numerical artifacts, all simulations are run for 1.5 FTTs at first to pass the transient process. After that, these simulations are run for another 1.5 FTTs to collect flow statistics. 

\begin{table}
\caption{Mesh parameters of the simulations for the flow over Boeing Gaussian bump at $Re_L=2\times10^6$.}
\begin{center}
\vspace{-0.5cm}
\begin{tabular}{l r c r r}
\toprule
Mesh & \multicolumn{1}{c}{$N_{\text{CV}}$} & \multicolumn{1}{c}{$\min \Delta_x/L$} & \multicolumn{1}{c}{$\min \Delta_y/L$} & \multicolumn{1}{c}{$\min \Delta_z/L$}\\
\midrule
Coarse & $6\times 10^6$ & $3.8\times10^{-3}$ & $2.6\times10^{-3}$ & $2.6\times10^{-3}$ \\
Medium & $37\times 10^6$ & $1.9\times10^{-3}$ & $1.3\times10^{-3}$ & $1.3\times10^{-3}$ \\
Fine   & $100\times 10^6$ & $1.3\times10^{-3}$ & $9\times10^{-4}$ & $9\times10^{-4}$ \\
\bottomrule
\end{tabular}
\label{tab:3d_grid}
\end{center}
\end{table}

\subsection{Results and Discussion}

\begin{figure}
\begin{center}
\includegraphics[width=0.95\textwidth]{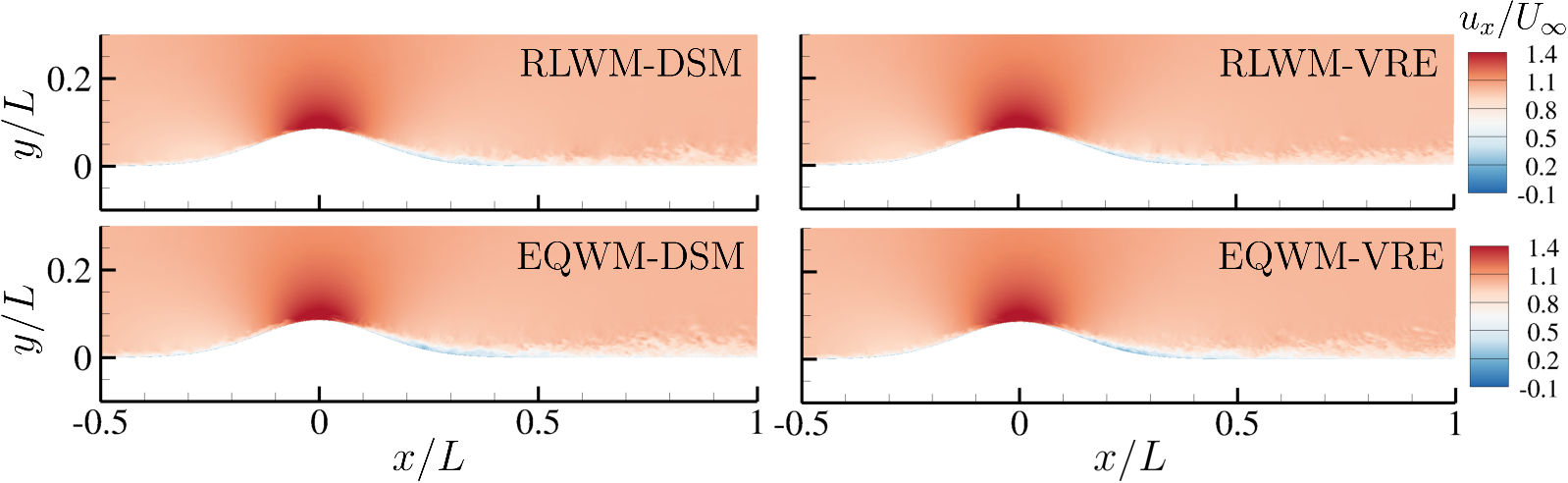}
\vspace{-0.3em}
\caption{Instantaneous streamwise velocity $u_x/U_\infty$ in an $x-y$ plane at $z/L=0$ from medium-mesh simulations.}
\label{fig:bump_xy_ux}
\end{center}
\end{figure}

Figure~\ref{fig:bump_xy_ux} shows contours of the instantaneous streamwise velocity $u_x/U_\infty$ in an $x$-$y$ plane at $z/L=0$ obtained from the medium-mesh simulations with different wall models and SGS models. The flow gradually accelerates on the windward side of the bump, and the velocity reaches its maximum value at the bump peak. Downstream of the peak, the flow decelerates over the leeward side of the bump, and the boundary layer thickens rapidly. Note that the flow is attached over the entire bump surface in the simulations, which is qualitatively different with the experimental observations that the flow is separated on the leeward side of the bump \citep{williams2020experimental,gray2021new,gray2022experimental,gray2022benchmark,gluzman2022simplified}. Detailed quantitative comparison of mean velocity from different simulations will be discussed later. 


\begin{figure}
\begin{center}
\includegraphics[width=1\textwidth]{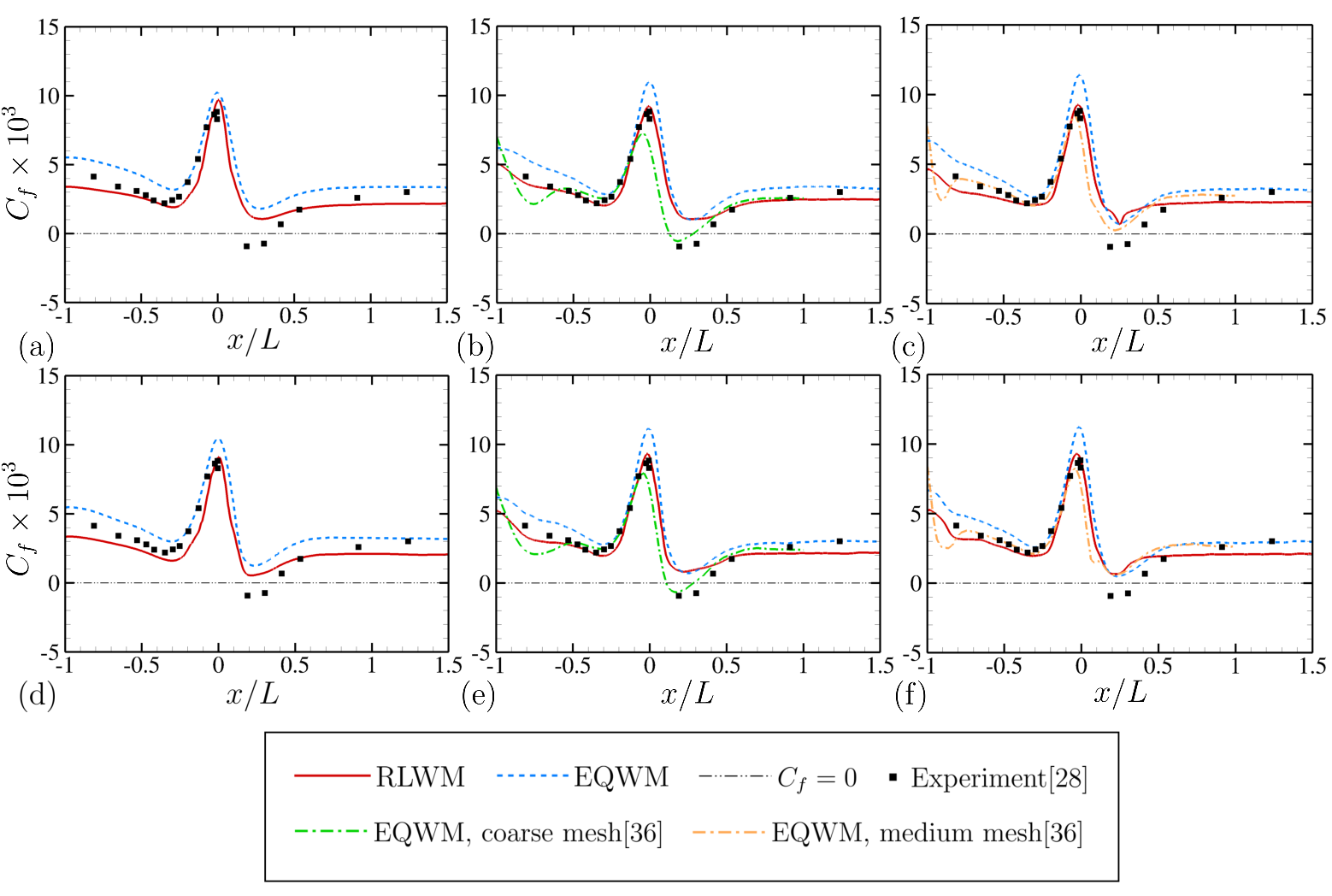}
\vspace{-1.6em}
\caption[]{The distributions of mean skin friction coefficient along the bump surface at $z/L=0$ from the simulations with (a,d) coarse mesh, (b,e) medium mesh and (c,f) fine mesh. The top row presents simulations using the DSM, while the bottom row features cases using the Vreman model.}
\label{fig:bump_cf}
\end{center}
\end{figure}

Additional quantities of interest for the bump-flow simulations include the mean skin friction coefficient, $C_{f}$, and the mean pressure coefficient, $C_p$, which are defined as
\begin{equation}\label{eq:qoi}
C_{f} = \frac{\overline{\tau_{w}}}{\frac{1}{2}\rho_\infty U_\infty^2} \qquad \textrm{and} \qquad C_p = \frac{\overline{p_w}-p_\infty}{\frac{1}{2}\rho_\infty U_\infty^2}~,
\end{equation}
respectively. Here, the mean pressure at the inlet near the top boundary is chosen as the reference freestream pressure $p_\infty$. Figure~\ref{fig:bump_cf} presents the distributions of the mean skin friction coefficient over the bump surface. For comparison, the data from the experiment using the same geometry at the same Reynolds number from \citet{gray2022experimental} is also shown. Additionally, the results from the coarse-mesh and medium-mesh EQWM simulations conducted by \citet{agrawal2022non} are depicted in the figure. It is important to note that the Reynolds number in their EQWM simulations is $Re_L=3.41\times10^6$. However, experimental measurements \citep{gray2022experimental} suggest that the variations in $C_f$ are not significant when $Re_L$ increases from $2\times10^6$ to $3.41\times10^6$. In their simulations, both the DSM \citep{germano1991dynamic,lilly1992} and the Vreman model ($c\approx0.07$) \citep{vreman2004eddy} are employed separately, with a plug flow inlet condition applied at $x/L=-1$, identical to the approach taken in our simulations. Moreover, isotropic Voronoi meshes are used. Within the boundary layer, their coarse mesh offers slightly better resolution compared to the medium mesh in our study, and their medium mesh is more refined than our fine mesh. Specifically, the size of their medium-mesh cell is $40\%$ smaller than that of our fine-mesh cell. From the results in Fig.~\ref{fig:bump_cf}, it can be noted that upstream of the bump, the $C_f$ results from these simulations exhibit a similar trend as the experiments. However, in the experiments, a large separation bubble appears downstream of the bump peak, which is not captured in all simulations. Considering the performance of different wall models, the simulations using the EQWM overpredict $C_f$ near the bump peak. In contrast, the developed RLWMs provide more accurate predictions for this region, and as the mesh resolutions increase, the predictions of $C_f$ are consistent and agree well with the experimental data. While the performance of the current RLWMs within the region upstream of the bump is better than that of the EQWM, it is still suboptimal. This is expected as the model training does not encompass laminar or transitional flows. On the leeward side of the bump, the predictions of $C_f$ from both wall models are comparable. However, further downstream, the developed RLWMs provide lower skin friction than the EQWM. Experimental studies \citep{gray2021new,gray2022experimental,gray2022benchmark,gluzman2022simplified} have shown that the friction Reynolds number, $Re_\tau$, increases rapidly downstream of the bump as the flow recovers from a non-equilibrium state. Specifically, $Re_\tau$ becomes larger than 7000 when $x/L\geq0.5$, and this $Re_\tau$ significantly exceeds what the RLWMs encountered during training. As detailed in Appendix II, the error in $C_f$ prediction by the RLWMs in flat-plate channel flow simulations grows with increasing $Re_\tau$. This trend potentially explains the suboptimal predictions of RLWMs in the downstream ZPG region of the bump flow. In addition, compared to the results from our simulations, the $C_f$ data from \citet{agrawal2022non} exhibit strong undulations ahead of the bump. More significantly, with mesh refinement, the flow separation observed in the coarse-mesh simulations disappears. This phenomenon is likely related to the behavior of SGS model \citep{agrawal2022non,zhou2024sensitivity}.

\begin{figure}
\begin{center}
\includegraphics[width=1\textwidth]{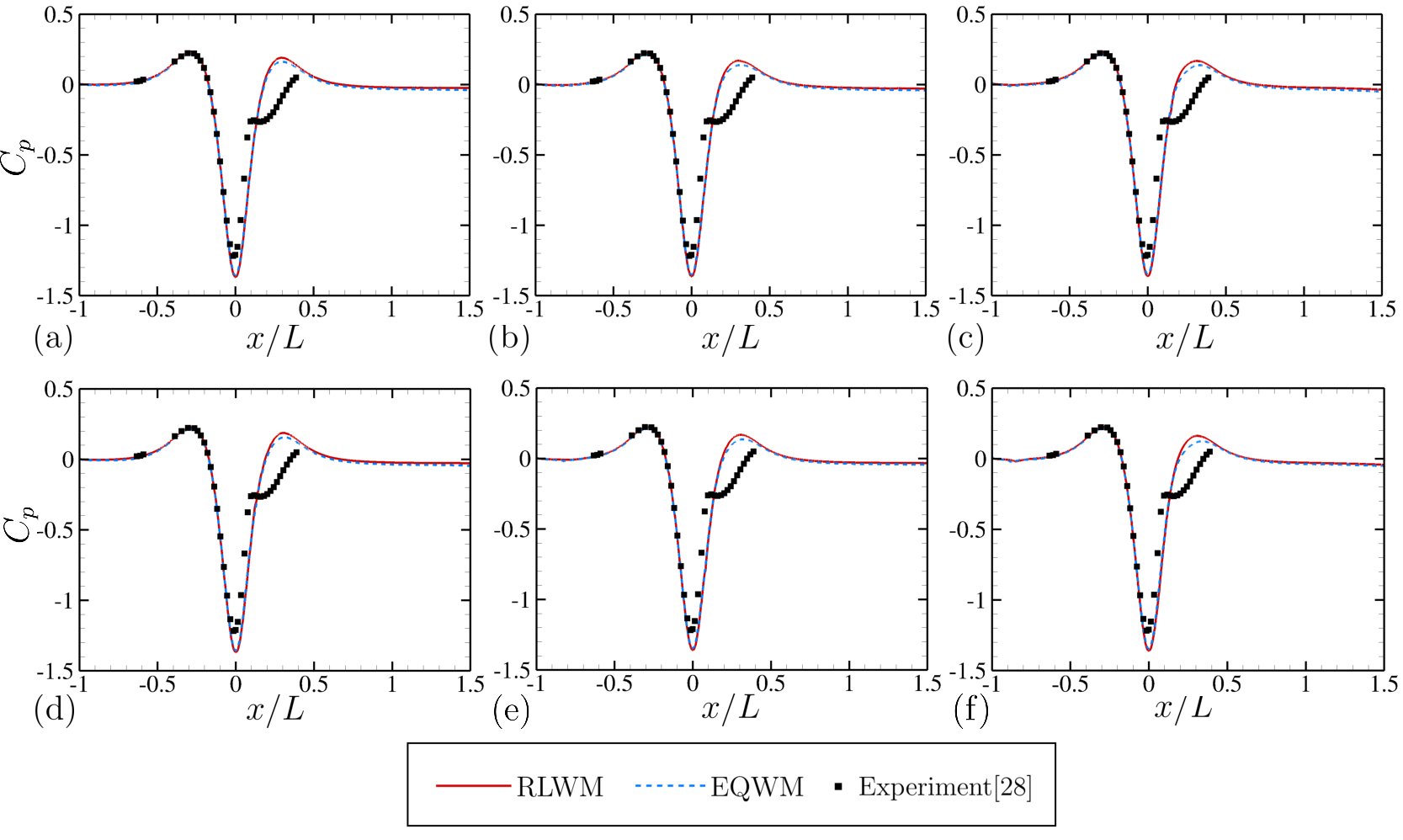}
\vspace{-1.6em}
\caption[]{The distributions of mean pressure coefficient along the bump surface at $z/L=0$ from the simulations with (a,d) coarse mesh, (b,e) medium mesh and (c,f) fine mesh. The top row presents simulations using the DSM, while the bottom row features cases using the Vreman model.}
\label{fig:bump_cp}
\end{center}
\end{figure}

Figure~\ref{fig:bump_cp} shows the distribution of $C_p$ along the bump surface at $z/L=0$. Results from all simulations with different computational meshes are included along with the experimental measurements of \citet{gray2022experimental} at the same Reynolds number. The $C_p$ distributions illustrate strong FPG immediately upstream of the bump peak. Downstream of the peak, the flow is first subjected to a very strong APG then followed by a mild FPG. Comparison of simulation results with experimental data reveals significant differences on the leeward side of the bump, which is expected due to the imprecise capture of the separation bubble in the simulations. Furthermore, the variations in mesh resolution do not significantly affect $C_p$ predictions in these simulations. In terms of wall model performance, the RLWMs and EQWM provide similar $C_p$ predictions, with only minor differences observed on the leeward side of the bump. In addition, when examining the $C_p$ distributions along the span, the simulation results align closely with each other and with the experimental data in the outer-span region ($z/L>0.2$). However, discrepancies become noticeable near the center of the bump, where a downstream separation bubble is expected. For brevity, detailed results are omitted here but can be found in \citep{zhou2023large}.

\begin{figure}
\begin{center}
\includegraphics[width=0.92\textwidth]{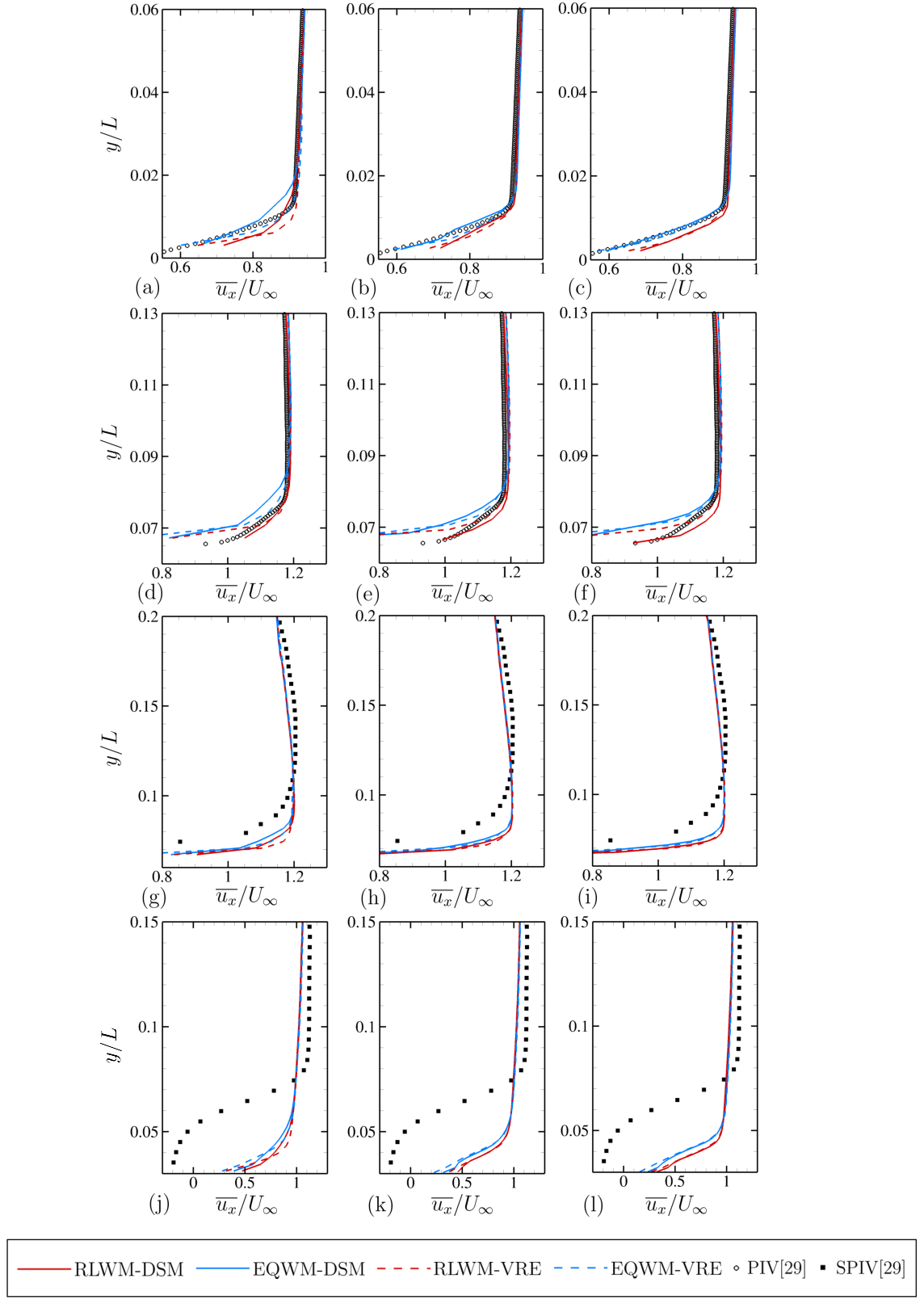}
\vspace{-0.5em}
\caption[]{Profiles of mean velocity $\overline{u_x}/U_\infty$ on an $x-y$ plane of $z/L=0$ at (a--c) $x/L = -0.4$, (d--f) $x/L=-0.1$, (g--i) $x/L=0.1$, and (j--l) $x/L=0.2$ from the simulations with coarse mesh (left), medium mesh (center), and fine mesh (right).}
\label{fig:bump_ux_stats}
\end{center}
\end{figure}

In Fig.~\ref{fig:bump_ux_stats}, profiles of the mean streamwise velocity are depicted at four streamwise stations $x/L=-0.4, -0.1, 0.1$ and 0.2, which are located within the regions with mild APG, strong FPG, and strong APG, respectively. These results quantitatively show the flow acceleration as well as deceleration and boundary-layer thickening on the bump surface. Based on the comparison in the figure, the simulation results downstream of the bump exhibit significant differences from the experiment conducted by \citet{gray2022benchmark}, with the simulations predicting a thinner TBL. This finding is consistent with the earlier observation of the instantaneous flow field. The simulations employing the two different RLWMs display significant differences at $x/L=-0.1$ where the FPG is pronounced. However, for most other locations, the results from both simulations align closely. When examining the EQWM simulations, velocity predictions exhibit small variation across different SGS models, and these predictions are comparable to those from the RLWM simulations on the leeward side of the bump. Nevertheless, at $x/L=-0.4$, the EQWM outperforms the RLWMs in velocity prediction, even though its prediction of mean skin friction is less ideal, as illustrated in Fig.~\ref{fig:bump_cf}.

While data-driven turbulence models often have limited applicability across different flow configurations \citep{duraisamy2019turbulence}, the current RLWMs, originally trained on the low-Reynolds-number periodic-hill channel flow, show promise in simulating the flow over the Boeing Gaussian bump, which has a different geometry as well as higher Reynolds number and is likely to exhibit different flow physics. Specifically, the RLWMs provide improved predictions for the skin friction near the peak of the bump and perform comparably to the EQWM with respect to the wall pressure and velocity field. For the sake of consistency, the SGS models used in the present bump-flow RLWM simulations correspond to those employed during model training. However, there is no strict requirement to utilize the same SGS model. In fact, previous simulations by \citet{zhou2023large}, which combined RLWM-DSM with the anisotropic minimum-dissipation SGS model \citep{rozema2015}, yielded improved predictions for the flow. In addition, a sensitivity analysis of WMLES by \citet{zhou2024sensitivity} highlighted the significant impact of SGS models on separated flow simulations. Specifically, this study revealed that the influence of wall boundary conditions on predicting separation bubble is overshadowed by that of the SGS models. Thus, the inability to capture separation in all the bump-flow simulations from this study can be largely attributed to the SGS models. 

In addition, comparison of the computational costs for different wall models in all test simulations reveals that the developed RLWMs consistently incur slightly higher costs than the EQWM. Specifically, the computational cost per time step for the RLWMs is approximately $7\%$ higher than that for the EQWM. It should be noted that the cost comparison focuses solely on evaluating the wall model and does not account for communication or flow solver costs.

\section{Conclusions}\label{sec:con}

Using MARL, we develop two wall models capable of adapting to varying pressure-gradient effects. These models are trained based on LES of low-Reynolds-number periodic-hill channel flow, each incorporating a different SGS model. Both models function as control policies for wall eddy viscosity, aiming to predict the correct wall-shear stress. During the training process, the optimized policy for each wall model is learned from LES with cooperating agents using the recovery of the correct wall-shear stress as a reward. The developed wall models are first validated in the LES of the periodic-hill configuration at the same Reynolds number of model training. The wall models provide good predictions of mean wall-shear stress, mean wall pressure, and the mean velocity as well as Reynolds stress in the flow field. The test results also show that the developed models outperform the EQWM. The performance of the developed models is further evaluated at two higher Reynolds numbers ($Re_H=19000$ and $37000$). The models achieve good predictions for the mean wall-shear stress, and show promising results for velocity statistics at $Re_H=19000$.

To further investigate the applicability and robustness of the developed wall models, simulations of flow over the Boeing Gaussian bump at a moderately high Reynolds number are conducted. The flow geometry is consistent with the experiments \citep{gray2021new,gray2022experimental,gray2022benchmark,gluzman2022simplified}, and the Reynolds number based on the freestream velocity and the width of the bump is $2\times10^6$. The results of mean skin friction and pressure on the bump surface and the velocity statistics of the flow field are compared to those from the EQWM simulations and published experimental data sets. It shows that the developed wall models successfully capture the acceleration and deceleration of the TBL on the bump surface. In addition, these models offer more accurate predictions of skin friction around the peak of the bump, and their performance in predicting wall pressure and the velocity field is comparable to that of the EQWM.

On the foundation of the test simulations across various flow configurations, it is evident that the RLWMs in this study, developed based on the training with low-Reynolds-number periodic-hill channel flow, have gained valuable insights into the key physics of complex flows with diverse pressure gradients. More importantly, this study underscores the promise of the MARL framework for sophisticated turbulence modeling. However, these test simulations also indicate that the performance of the developed RLWMs is not yet optimal, particularly at higher Reynolds numbers, and these models show sensitivity to mesh resolution. It is important to highlight that the RLWMs were trained \emph{in situ} with WMLES for periodic-hill channel flow at a relatively low Reynolds number ($Re_H=10595$), without specific optimization of model components, including environmental states, rewards, and actions. Consequently, further enhancements in model performance and robustness could potentially be realized by broadening the training across a wider range of Reynolds numbers and flow geometries, or through meticulous optimization of the model components. Moreover, the role of SGS models in shaping the wall model development is also discernible, especially given their significant influence in simulations of separated turbulent flow \citep{zhou2024sensitivity}. To improve the reliability and versatility of RL-based wall model in wider applications, a potential avenue worth exploring is extending the wall model to become a unified model that encompasses both SGS and wall boundary modeling, as seen in models like the building-block flow model \citep{ling2022wall,arranz2023wall,lozano2023building}.  

\section*{Appendix I: Convergence of Model Training} \label{sec:append_train}
\begin{figure}
\begin{center}
\includegraphics[width=0.5\textwidth]{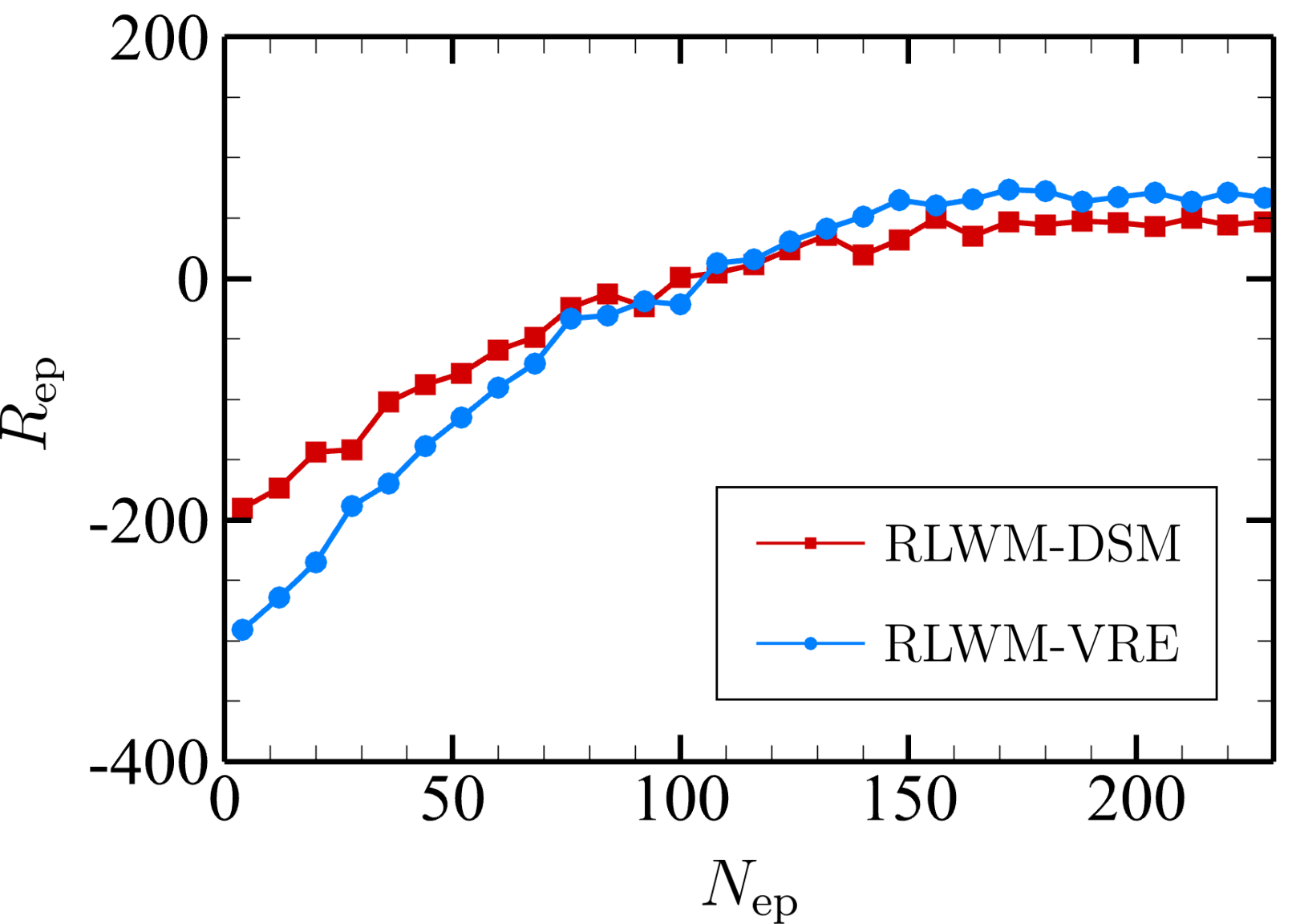}
\vspace{-0.5em}
\caption[]{Averaged episode reward $R_{\text{ep}}$ against episode number $N_{\text{ep}}$ during the training of RLWMs.}
\label{fig:RL_train}
\end{center}
\end{figure}

As detailed in Section~\ref{sec:training}, the training of models is based on a series of periodic-hill channel flow simulations. In order to accelerate the training, 8 simulations (episodes) were run in parallel. Subsequently, the policies of the trained models are updated after every set of 8 episodes, defining one iteration. Figure~\ref{fig:RL_train} shows the averaged episode rewards $R_{\text{ep}}$ for all iterations during the training of the aforementioned two RLWMs with different SGS models. The averaged episode reward for each iteration is defined as $R_{\text{ep}}=(1/8)\sum_{l=1}^{8}\sum_{k=1}^{N_{\text{k}}}\sum_{j=1}^{N_j} r_{j,k,l}$, where $j$ denotes the $j$-th sampling step in one episode, $N_j$ represents the total number of sampling steps in one episode, $k$ denotes the index of RL agent, $N_k$ is the total count of RL agents in each episode and $l$ marks the index of episode in one iteration. From the presented results, it is evident that the training converges around 200 episodes or 25 iterations for both cases.

\section*{Appendix II: Testing for flat-plate channel flow}\label{sec:append_channel}
\begin{figure}
\begin{center}
\includegraphics[width=\textwidth]{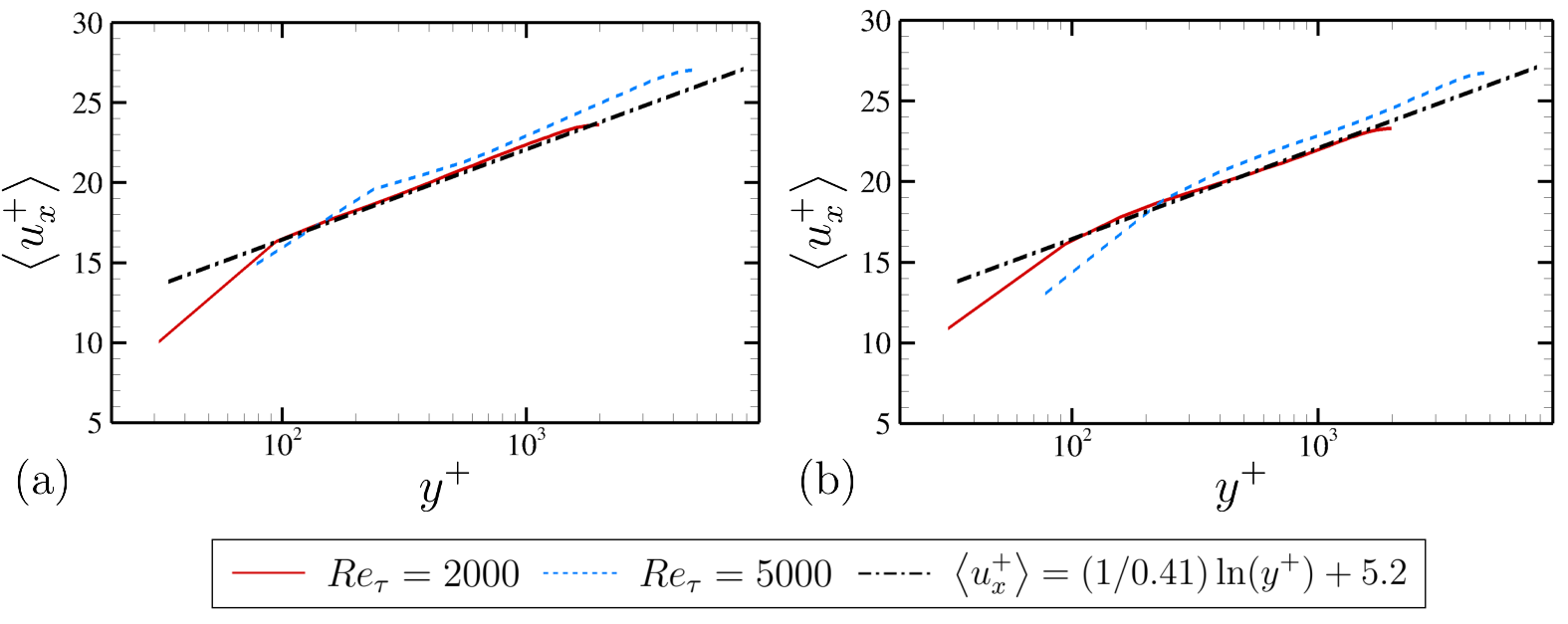}
\vspace{-1.3em}
\caption[]{Inner-scaled mean streamwise velocity profiles from flat-plate channel flow simulations with (a) RLWM-DSM and (b) RLWM-VRE at two different Reynolds numbers.}
\label{channel_vel}
\end{center}
\end{figure}

To validate the trained models and assess the influence of LLM, test simulations of flat-plate channel flows are conducted at two distinct friction Reynolds numbers, $Re_\tau=2000$ and $Re_\tau=5000$. The computational domain dimensions are set to $10H_c \times 2H_c \times 3H_c$ in the streamwise ($x$), wall-normal ($y$), and spanwise ($z$) directions, respectively, with $H_c$ representing the half-channel height. This domain is discretized uniformly into 200, 64, and 60 cells in the $x$, $y$, and $z$ directions, respectively. The flow is driven by a constant pressure gradient in the $x$ direction. Periodic boundary conditions are applied at both ends of the domain in the streamwise and spanwise directions, while the bottom and top boundaries employ the previously developed RLWM-DSM and RLWM-VRE. The number of RL agents positioned above the bottom and top walls matches the number of wall cells, with each agent situated at the upper surface of the wall-adjacent mesh cells. The effective wall eddy viscosity, $\nu_{t,w}$, is dynamically updated at each time step in response to local instantaneous flow states. To ensure consistency across simulations, both the DSM \citep{germano1991dynamic,lilly1992} and the Vreman model \citep{vreman2004eddy} are utilized in separate simulations. A maximum CFL number of 1 is employed throughout all simulations. Initially, each simulation run for 100 FTTs to pass the initial transient phase. Subsequently, an additional 100 FTTs are executed to collect flow statistics. We first assessed the capability of RLWMs to predict skin friction coefficient at the wall. Table~\ref{tab:error} presents the errors in mean skin friction coefficient obtained from the RLWM simulations, which exhibits an increase in error with Reynolds number. Additionally, Fig.~\ref{channel_vel} displays the inner-scaled mean streamwise velocity profiles. Overall, the RLWMs demonstrate reasonable predictions for flat-plate channel flows, but LLM becomes apparent at higher $Re_\tau$, aligning with the observed trends in skin friction errors. It is important to note that the RLWMs are originally trained using periodic-channel flow simulations characterized by spatially and temporally varying pressure gradients. In these training scenarios, the maximum $Re_\tau$ is approximately 1900, lower than that of the flat-plate channel flow cases examined here.


\section*{Acknowledgments}
This work was supported by National Science Foundation (NSF) grant No.~2152705 and the Stanford University Center for Turbulence Research Summer Program. Computer time was provided by the Discover project at Pittsburgh Supercomputing Center through allocation PHY230012 from the Advanced Cyberinfrastructure Coordination Ecosystem: Services \& Support (ACCESS) program, which is supported by NSF grants No.~2138259, No.~2138286, No.~2138307, No.~2137603, and No.~2138296. The authors gratefully acknowledge Dr. Kevin P. Griffin and Michael P. Whitmore for valuable discussions.

\begin{table}
\caption{Errors in mean skin friction coefficient obtained from flat-plate channel flow simulations with RLWM-DSM and RLWM-VRE at two different Reynolds numbers.}
\begin{center}
\vspace{-0.5cm}
\begin{tabular}{l c c}
\toprule
$Re_\tau$ & Error in $C_f$ for RLWM-DSM & Error in $C_f$ for RLWM-VRE \\
\midrule
2000 & $-1.2\%$ & $-1.6\%$ \\
5000 & $-7.9\%$ & $-7.1\%$ \\
\bottomrule
\end{tabular}
\label{tab:error}
\end{center}
\end{table}

\end{document}